\documentclass[aps,prd,superscriptaddress,twoside,twocolumn,nofootinbib,showpacs,
floatfix]{revtex4-2}
\usepackage{amsmath,amssymb}
\usepackage{graphicx,bm}
\usepackage{slashed,color}
\usepackage[colorlinks=true,pdfstartview=FitV,bookmarks=true,bookmarksnumbered=true,
breaklinks]{hyperref}

\hypersetup{linkcolor=red, citecolor=blue, urlcolor=blue}

\usepackage[normalem]{ulem} 
\usepackage[dvipsnames]{xcolor} 
\begin{document}
\title{\boldmath Production of $\phi \Lambda$, $D_s^{*-} \Lambda_c^+$, and $J/\psi
\Lambda$ in kaon-induced reactions off the nucleon}
\author{Sang-Ho Kim}
\email{shkimphy@gmail.com}
\affiliation{Department of Physics and Origin of Matter and Evolution of Galaxies
(OMEG) Institute, Soongsil University, Seoul 06978, South Korea}

\date{\today}

\begin{abstract}

We investigate the reaction mechanism of strangeness production in $K^- p \to \phi
\Lambda$ within a hybrid Regge approach, taking into account $t$-channel $K$- and
$K^*$-Reggeon exchanges.
We present results for the total cross section, $t$-dependent differential cross
sections, and spin-density matrix elements (SDMEs), and compare them with the
available experimental data.
We find that $K^*$-Reggeon exchange provides the dominant contribution, while
$K$-Reggeon exchange remains nonnegligible, particularly in describing the SDMEs.
In contrast, the $s$-channel $\Lambda$ and $u$-channel nucleon exchanges are
negligible.
To obtain reliable predictions for the open-charm reaction $K^- p \to D_s^{*-}
\Lambda_c^+$, we employ a Quark–Gluon String Model (QGSM)-motivated Regge framework
that incorporates both pseudoscalar- and vector-Reggeon exchanges.
Within this framework, the Regge trajectories $\alpha(t)$ and energy-scale
parameters $s_0$ are determined consistently, thereby constaining the model and
reducing theoretical ambiguities.
For the hidden-charm reaction $K^- p \to J/\psi \Lambda$, we use the same
quark-level diagrammatic correspondence.
The total cross sections for $K^- p \to D_s^{*-} \Lambda_c^+$ and $K^- p \to J/\psi
\Lambda$ are suppressed by approximately 5–6 and 8–9 orders of magnitude,
respectively, compared with that for $K^- p \to \phi \Lambda$.
We also examine possible $s$-channel contributions from the hidden-charm pentaquark
states with strangeness, $P_{cs}(4337)^0$ and $P_{cs}(4459)^0$, to both $D_s^{*-}
\Lambda_c^+$ and $J/\psi \Lambda$ production.

\end{abstract}

\maketitle
\section{Introduction}

Open- and hidden-charm production constitute central topics in current hadron-physics
programs at the $\rm{{\bar P}ANDA}$ experiment at FAIR (Facility for Antiproton and
Ion Research)~\cite{PANDA:2009yku} and at J-PARC (Japan Proton Accelerator Research
Complex)~\cite{Aoki:2021cqa}.
At $\rm{{\bar P}ANDA}$, the high-intensity antiproton beam with momenta up to 15
GeV/$c$ will enable systematic investigations of charm dynamics covering
open-charm processes such as $\bar p p \to D \bar D^{(*)}$, $\Lambda_c \bar
\Lambda_c^{(*)}$, $\Sigma_c \bar \Lambda_c^{(*)}$, as well as hidden-charm production
channels like $\bar p p \to J/\psi X$, $\eta_c \gamma$~\cite{Andreotti:2005vu,
PANDA:2016scz}.
Corresponding theoretical studies have been performed within various frameworks,
including the J\"ulich meson-baryon coupled-channel model~\cite{Haidenbauer:2014rva,
Haidenbauer:2016pva}, effective Lagrangian and Regge frameworks~\cite{Titov:2008yf,
Shyam:2014dia,Sangkhakrit:2020wyi}, NRQCD factorization~\cite{Bodwin:1994jh}, and
hadronic pole models~\cite{Barnes:2006ck,Lundborg:2005am}.

In parallel, J-PARC has developed complementary programs employing meson beams.
The E50 spectrometer, for example, is designed to detect charged open-charm hadrons
produced in $\pi^- p \to D^{*-} Y_c^{*+}$ at an incident pion momentum of 20 GeV/$c$,
where $Y_c^{*+}$ denotes excited charmed baryons, such $\Lambda_c^*$'s and
$\Sigma_c^*$'s~\cite{J-PARC:P50}.
With its wide angular coverage and high momentum resolution, this experiment is
expected to provide valuable information on charmed-baryon spectroscopy~\cite{
Kim:2014qha,Kim:2015ita,Shim:2019yxn}.
More recently, a new proposal (P111) has also been submitted to study hidden-charm
production near threshold via $\pi^- p \to J/\psi n$~\cite{J-PARC:P111}.
Existing theoretical predictions for this cross section span orders of
magnitude, ranging from $\sigma_{\rm tot}$ $\sim 0.1$ pb~\cite{Kim:2016cxr} to $\sim
50$ pb~\cite{Sibirtsev:1998cs} and $\sim 1.0$ nb~\cite{Wu:2013xma} depending 
strongly on the assumed mechanism and model inputs.
Reliable and model-constrained predictions are therefore essential for guiding
experimental feasibility and optimizing detector design.

In previous works, the author and collaborators investigated open-strangeness
($\pi^- p \to K^{(*)} \Lambda$) and open-charm ($\pi^- p \to D^{(*)} \Lambda_c^+$)
reactions~\cite{Kim:2015ita,Kim:2016imp,Kim:2017hhm} within a unified framework
based on the quark-gluon string model (QGSM)~\cite{Kaidalov:1982bq,Boreskov:1983bu,
Kaidalov:1986zs, Kaidalov:1994mda}.
In this picture, the annihilation of a $q \bar q$ pair in the initial state leads
to the formation of an intermediate string, which subsequently fragments into the
observed hadrons via planar diagrams.
A key advantage of the QGSM-based approach is that the two essential ingredients,
Regge trajectories $\alpha (t)$ and energy-scale parameters $s_0$, can be determined
consistently across both the strangeness and charm sectors.
Using the similarity of quark diagrams, hidden-strangeness ($\pi^- p \to \phi n$) and
hidden-charm ($\pi^- p \to J/\psi n$) reactions were also studied in a coherent
manner~\cite{Kim:2016cxr}.

Beyond pion-induced reactions, J-PARC has established an extensive program of
kaon-induced reactions using the high-intensity K1.8 beam line~\cite{Aoki:2021cqa,
Takahashi:2012cka,Agari:2012gj}.
Motivated by these developments, in the present work, we extend the QGSM-motivated
Regge framework~\cite{Kaidalov:1982bq,Boreskov:1983bu,Kaidalov:1986zs,
Kaidalov:1994mda} to kaon-induced reactions off the nucleon.
We first analyze the strangeness production process $K^- p \to \phi \Lambda$,
including both pseudoscalar $K$- and vector $K^*$-Reggeon exchanges in the $t$
channel.
We then provide predictions for the open-charm reaction $K^- p \to D_s^{*-}
\Lambda_c^+$ based on the same quark-level diagrammatic correspondence, where the
parameters associated with $D$- and $D^*$-Reggeon exchanges are constrained within
the QGSM framework.
For the hidden-charm reaction $K^- p \to J/\psi \Lambda$, we employ the same $K$-
and $K^*$-Reggeon exchange mechanisms but with appropriate coupling constants.
All the cutoff masses in the hadronic form factors are fixed by fitting the
available $K^- p \to \phi \Lambda$ data~\cite{Lindsey:1966zz,Ayres:1974aj,
Aguilar-Benitez:1972ngz}, therby reducing model ambiguities and allowing more
reliable predictions for charm production.
We present results for the total cross sections, $t$-dependent differential cross
sections, and spin-density matrix elements (SDMEs).
In particular, SDMEs encode the helicity structure of the reaction amplitude and
thus provide a stringent test of the underlying production mechanism~\cite{
Kim:2017hhm}.

The $K^- p \to J/\psi \Lambda$ reaction is especially interesting because it offers
direct access to possible hidden-charm pentaquark states with strangeness, $P_{cs}$,
in the $s$-channel~\cite{LHCb:2020jpq,LHCb:2021chn,LHCb:2022ogu}.
Recently, the LHCb Collaboration reported hidden-charm pentaquark candidates with
quark content $udsc\bar c$.
The $P_{cs}(4459)^0$ state was observed in the analysis of the $\Xi_b^- \to J/\psi
\Lambda K^-$ decay~\cite{LHCb:2020jpq}, while the $P_{cs}(4337)^0$ state was found in
the $B^- \to J/\psi \Lambda \bar p$ decay~\cite{LHCb:2022ogu}.
Their measured masses and widths are
\begin{align}
M_{P_{cs}} = &\, 4458.8 \pm 2.9_{-1.1}^{+4.7},\,
\Gamma_{P_{cs}} = 17.3 \pm 6.5_{-5.7}^{+8.0}~\cite{LHCb:2020jpq},
\cr
M_{P_{cs}} = &\, 4338.2 \pm 0.7 \pm 0.4,\,
\Gamma_{P_{cs}} = 7.0 \pm 1.2 \pm 1.3~\cite{LHCb:2022ogu},
\label{eq:Pentaquark}
\end{align}
respectively, in units of MeV.
The dominant open-charm decay modes of these $P_{cs}$ states have been studied
theoretically within meson-baryon molecular pictures~\cite{Wang:2019nvm,
Peng:2020hql,Chen:2020kco,Yan:2022wuz,Wang:2022mxy,Wang:2022nqs,Zhu:2022wpi,
Feijoo:2022rxf,Wang:2023ael,Wang:2023eng,Roca:2024nsi,Yang:2022ezl}, some of which
predict a sizable (possibly dominant) branching ratio into the $D_s^* \Lambda_c$
channel~\cite{Chen:2021tip,Xiao:2021rgp}.
This suggests that the $P_{cs} (4459)^0$ state may also be probed through the
associated-production reaction $K^- p \to D_s^{*-} \Lambda_c^+$.
Accordingly, we investigate possible $s$-channel contributions from the $P_{cs}$
states to both the $K^- p \to D_s^{*-}\Lambda_c^+$ and $K^- p \to J/\psi \Lambda$
reactions.

This paper is organized as follows.
In Sec.~\ref{Sec:II}, we present the formalism of the hybrid Regge approach and
specify the model ingredients, including coupling constants, Regge trajectories,
energy-scale parameters, and form factors.
Section~\ref{Sec:III} contains numerical results for the total and differential cross
sections as well as SDMEs, followed by discussion.
Finally, Sec.~\ref{Sec:IV} summarizes our findings and provides concluding remarks.

\section{Theoretical Framework}
\label{Sec:II}

\begin{figure}[h]
\centering
\includegraphics[scale=0.43]{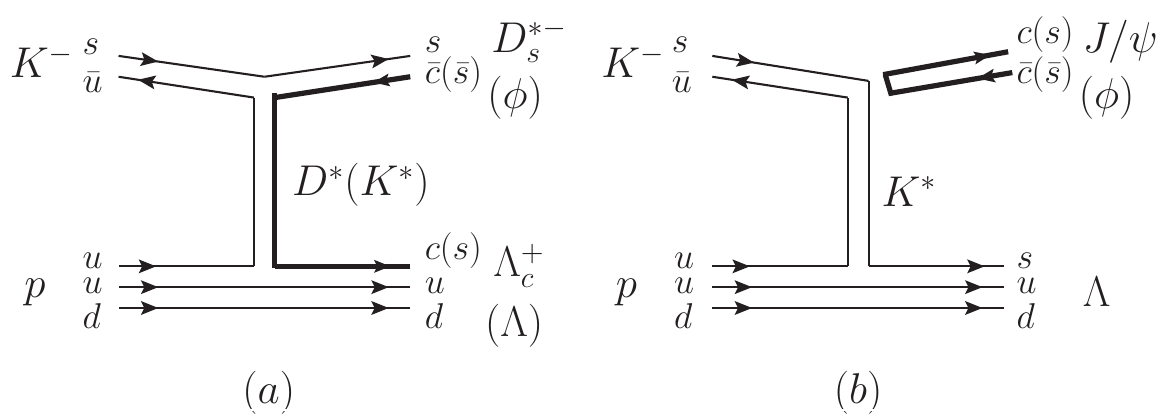}
\caption{Quark diagrams for (a) $K^- p \to \phi \Lambda, D_s^{*-} \Lambda_c^+$ and
(b) $K^- p \to \phi \Lambda, J/\psi \Lambda$ in the $t$ channel.}
\label{FIG01}
\end{figure}

In this section, we introduce a hybrid Regge model that combines an effective
Lagrangian approach with Regge phenomenology.
As illustrated in Fig.~\ref{FIG01}, the strangeness reaction $K^- p \to \phi
\Lambda$ can be described by two quark diagrams.
In Fig.~\ref{FIG01}(a), replacing the strange quark in the propagator with a charm
quark ($s \to c$) leads to the open-charm process $K^- p \to D_s^{*-} \Lambda_c^+$.
In Fig.~\ref{FIG01}(b), the corresponding replacement in the produced vector meson
results in the hidden-charm process $K^- p \to J/\psi \Lambda$.

\begin{figure}[h]
\centering
\includegraphics[scale=0.55]{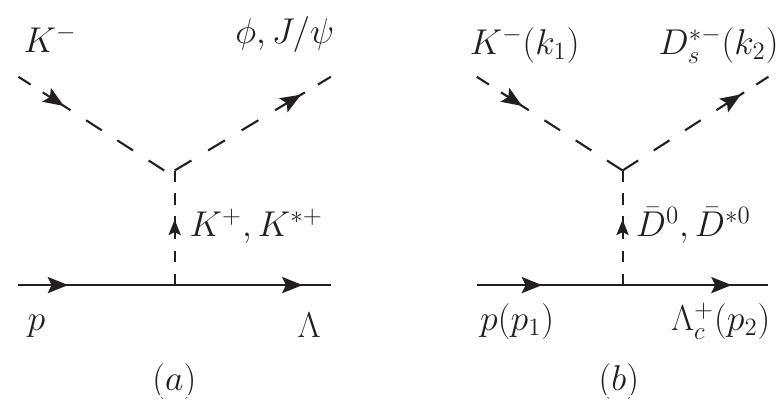}
\caption{Feynman diagrams for (a) $t$-channel $K$- and $K^*$-Reggeon exchanges for
$K^- p \to \phi \Lambda, J/\psi \Lambda$, and
(b) $t$-channel $D$- and $D^*$-Reggeon exchanges for $K^- p \to D_s^{*-} \Lambda_c^+$.}
\label{FIG02}
\end{figure}

For the $K^- p \to \phi \Lambda$ and $K^- p \to J/\psi \Lambda$ reactions, the
$t$-channel contributions are described by  pseudoscalar $K$- and vector $K^*$-Reggeon
exchanges [Fig.~\ref{FIG02}(a)].
For $K^- p \to D_s^{*-} \Lambda_c^+$, we include pseudoscalar $D$- and vector
$D^*$-Reggeon exchanges in the $t$ channel [Fig.~\ref{FIG02}(b)].

\subsection{Strangeness production: $K^- p \to \phi \Lambda$}
\label{Sec:II-1}

We begin with the $t$-channel process of the $K^- p \to \phi \Lambda$ reaction shown
in Fig.~\ref{FIG02}.
The four momenta of the initial $K$ and proton are denoted by $k_1$ and $p_1$,
respectively, while those of the final $\phi$ and $\Lambda$ are denoted by $k_2$ and
$p_2$.
The effective Lagrangians for the $\phi K K$ and $\phi K^* K$ vertices are given by
\begin{align}
\mathcal L_{\phi K K} = &\,
-ig_{\phi K K} ( K^- \partial_\mu K^+ - \partial_\mu K^- K^+ ) \phi^\mu,
\cr
\mathcal L_{\phi K^* K} = &\,
{g_{\phi K^* K}} \varepsilon^{\mu\nu\alpha\beta}
\partial_\mu \phi_\nu (\partial_\alpha K^{*-}_\beta K^+ + \partial_\alpha K^{*+}_\beta K^-),
\cr
\label{eq:Lag1}
\end{align}
where $K$, and $K^*$, and $\phi$ represent the fields of $K (494,0^-)$,
$K^* (892,1^-)$, and $\phi (1020,1^-)$ mesons, respectively.
The coupling constant $g_{\phi K K} = 4.48$ is fixed from the branching ratio
${\mathcal B} (\phi \to K^+ K^-) = 49.9\,\%$~\cite{PDG:2024cfk} using the width
of $\Gamma_\phi =$ 4.249 MeV and the relation,
\begin{align}
\Gamma (\phi \to K^+ K^-) = \frac{p_K^3}{6 \pi M_\phi^2} g_{\phi K K}^2,
\label{eq:DWidth}
\end{align}
where $p_K = \sqrt{M_\phi^2 - 4 M_K^2} / 2$.
To obtain the coupling constant $g_{\phi K^* K}$, we use the SU(3) flavor symmetry
relation,
\begin{align}
g_{\phi K^* K} = g_{\rho \omega \pi} / \sqrt2,
\end{align}
where the coupling $g_{\rho \omega \pi}$ is taken from the hidden gauge approach,
\begin{align}
g_{\rho \omega \pi} = \frac{N_c g_{\phi \pi \pi}^2}{8 \pi^2 f_\pi} = 14.4 \,\rm{GeV}^{-1},
\end{align}
with $N_c = 3$, $f_\pi = 93$ MeV, and $g_{\rho \pi \pi} = 5.94$~\cite{Oh:2004wp,
Kim:2023jij}.
$g_{\rho \pi \pi}$ is calculated from the branching ratio $\mathcal{B}
(\rho \to \pi \pi) \sim$ 1~\cite{PDG:2024cfk}.

The effective Lagrangians for the meson-baryon octet vertices are written as
\begin{align}
\mathcal L_{K N \Lambda} =&\,
\frac{g_{K N \Lambda}}{M_N+M_\Lambda} \bar N \gamma_\mu  
\gamma_5 \Lambda \partial^\mu K + {\rm H.c.}.
\cr
\mathcal L_{K^* N \Lambda} = &\,
-g_{K^* N \Lambda} \bar N \left[ \gamma_\mu \Lambda - 
\frac{\kappa_{K^* N \Lambda}}{2M_N} \sigma_{\mu\nu} \Lambda 
\partial^\nu \right] K^{*\mu} + \mathrm{H.c.},
\cr
\label{eq:Lag2}
\end{align}
where $N$ and $\Lambda$ stand for the nucleon and $\Lambda(1116)$ baryon fields,
respectively.
The coupling constants are taken from the Nijmegen soft-core model (NSC97a)~\cite{
Rijken:1998yy,Stoks:1999bz},
\begin{align}
g_{K N \Lambda} = & -13.4,
\cr
g_{K^* N \Lambda} = & -4.26, \,\,\, \kappa_{K^* N \Lambda} = 2.66.
\label{eq:Coupl1}
\end{align}

The $t$-channel Regge amplitudes are constructed within a hybrid approach by
replacing the Feynman propagators for pseudoscalar $K$- and vector $K^*$-meson
exchanges with the corresponding Regge propagators associated with their
Regge trajectories~\cite{Kim:2015ita,Titov:2008yf},
\begin{align}
T_K  (s,t) = &\, \mathcal M_K(s,t)
\left( \frac{s}{s_K^{K N : \phi \Lambda}} \right)^{\alpha_K(t)}
\cr
& \times \Gamma (-\alpha_K(t)) \alpha_K' F_{PS}^2(t),
\cr
T_{K^*} (s,t) = &\, \mathcal M_{K^*}(s,t)
\left( \frac{s}{s_{K^*}^{K N : \phi \Lambda}} \right)^{\alpha_{K^*}(t)-1}
\cr
& \times \Gamma (1-\alpha_{K^*}(t)) \alpha_{K^*}' F_V^2(t),
\label{eq:RegAmpl1}
\end{align}
where the amplitudes $\mathcal M_K$ and $\mathcal M_{K^*}$ are derived from the
Lagrangians in Eq.~(\ref{eq:Lag1}), respectively,
\begin{align}
\mathcal M_{K}^\mu = &\,2 i
\frac{g_{\phi K K} g_{K N \Lambda}}{M_N + M_\Lambda}
\gamma_\nu \gamma_5 k_1^\mu (k_2 - k_1)^\nu,
\cr
\mathcal M_{K^*}^\mu = &\,
g_{\phi K^* K} g_{K^* N \Lambda}
\epsilon^{\mu\nu\alpha\beta}
\cr & \times
\left [ \gamma_\nu - \frac{i\kappa_{K^* N \Lambda}}{2 M_N} 
\sigma_{\nu \lambda}(k_2-k_1)^\lambda \right ] k_{2 \alpha} k_{1 \beta},
\label{eq:Ampl1}
\end{align}
with $\mathcal M = \varepsilon_\mu^* \bar{u}_\Lambda \, \mathcal M^\mu \,u_N$.
Here $u_N$ and $u_\Lambda $ denote the Dirac spinors of the initial nucleon and
the final $\Lambda$, respectively, and are normalized to $\bar u_B u_B =1$.
The four-vector $\varepsilon_\mu$ stands for the polarization of the outgoing
$\phi$ meson.

The form factor in Eq.~(\ref{eq:RegAmpl1}) is introduced to dress the vertices in the
diagrams.
We adopt the following form:
\begin{align}
F_{PS(V)}(t) = \frac{\Lambda_{PS(V)}^2}{\Lambda_{PS(V)}^2 - t}.
\label{eq:FormFac_t}
\end{align}
The differential cross section $d\sigma/dt$ is expressed as 
\begin{align}
\frac{d\sigma}{dt} = \frac{M_N M_\Lambda}{16 \pi (p_{\rm c.m.})^2 s}
\frac{1}{2}\sum_{\lambda_V,s_f,s_i}|T|^2,
\label{eq:Def:dsdt}
\end{align}
where $p_{\mathrm{c.m.}}$ denotes the kaon momentum in the center-of-mass (c.m.) frame.
$s_i$, $s_f$, and $\lambda_V$ label the helicity states of the nucleon, the
$\Lambda$ baryon, and the $\phi$ meson, respectively.

$d\sigma/dt$ satisfies the following asymptotic behavior:
\begin{align}
\frac{d\sigma}{dt}(s \to \infty, t \to 0) \propto s^{2\alpha(0)-2}.
\label{eq:Asym:dsdt}
\end{align}
More specifically, the asymptotic behaviors of the squared amplitudes of
Eq.~(\ref{eq:Ampl1}) are derived as
\begin{align}
& \lim_{s \to \infty} \sum_{\lambda_V,s_f,s_i} |\mathcal M_K(s,t)|^2 \propto A,
\cr
& \lim_{s \to \infty} \sum_{\lambda_V,s_f,s_i} |\mathcal M_{K^*}(s,t)|^2 \propto - B s^2 t
+ C s^2 t^2,
\label{eq:Asym:Ampl}
\end{align}
where $A$, $B$, and $C$ are kinematics-independent constants, leading to distinct
cross-section shapes at very forward angles.
For $K$-Reggeon exchange, the cross section increases monotonically in this region,
whereas for $K^*$-Reggeon exchange the presence of an antisymmetric tensor in the
amplitude substantially reduces the cross section at very forward angles.

To determine the $K$-and $K^*$-Regge trajectories in Eq.~(\ref{eq:RegAmpl1}), we 
follow Ref.~\cite{Brisudova:1999ut}, where the so-called ``square-root'' trajectory is
adopted,
\begin{align}
\alpha(t) = \alpha(0) + \gamma [ \sqrt{T} - \sqrt{T-t} ],
\label{eq:SRTraj}
\end{align}
with $\gamma$ the universal slope and $T$ a scale parameter specific to each
trajectory.
In the limit $-t \ll T$, Eq.~(\ref{eq:SRTraj}) can be approximated by a linear form,
\begin{align}
\alpha(t) = \alpha(0) + \alpha' t,
\label{eq:LinTraj}
\end{align}
with the slope $\alpha'=\gamma/(2\sqrt{T})$.
In Ref.~\cite{Brisudova:1999ut}, the parameters $\sqrt{T}$ for the $\pi$- and
$\rho$-Regge trajectories were determined as
\begin{align}
\sqrt{T_\pi}  = &\, 2.82 \pm 0.05 \, {\rm GeV},
\cr
\sqrt{T_\rho} = &\, 2.46 \pm 0.03 \, {\rm GeV},
\label{UnivPara}
\end{align}
with $\gamma = 3.65 \pm 0.05 \, {\rm GeV^{-1}}$.
Following the same procedure, we extract the corresponding values of $\sqrt{T}$
for the $K$- and $K^*$-Regge trajectories.

The $\eta_s$- and $\phi$-Regge trajectories are determined using the additivity
relations of the intercepts and inverse slopes,
\begin{align}
2\alpha_{\bar s q}(0) = &\, \alpha_{\bar{q}q}(0) + \alpha_{\bar{s}s}(0),
\cr
2/\alpha'_{\bar s q} = &\, 1/\alpha'_{\bar q q} + 1/\alpha'_{\bar s s},
\label{eq:TrajRela1}
\end{align}
where the $\alpha_{\bar q q}(t)$, $\alpha_{\bar s q}(t)$, and $\alpha_{\bar s s}(t)$
denote the trajectories corresponding to $\pi$, $K$, and $\eta_s$ for pseudoscalar
mesons, and to $\rho$, $K^*$, and $\phi$ for vector mesons, respectively.
All Regge-trajectory parameters are summarized in Table~\ref{TAB:1}.
\begin{table}[h]
\begin{tabular}{cccc}
\hline\hline
&$\alpha(0)$
&\hspace{1.1em}$\sqrt{T}\,[{\rm GeV}]$\hspace{1.1em}
&$\alpha'\,[{\rm GeV}^{-2}]$
\\\hline
$\bar{q}q(\pi)$     \hspace{5.0em}& -0.0118 & 2.82 & 0.647   \\
$\bar{s}q(K)$       \hspace{5.0em}& -0.151  & 2.96 & 0.617   \\
$\bar{s}s(\eta_s)$  \hspace{5.0em}& -0.291  & 3.10 & 0.589   \\
\hline
$\bar{q}q(\rho)$ \hspace{5.0em}& 0.55  & 2.46 & 0.742   \\
$\bar{s}q(K^*)$  \hspace{5.0em}& 0.414 & 2.58 & 0.707   \\
$\bar{s}s(\phi)$ \hspace{5.0em}& 0.27  & 2.70 & 0.676   \\
\hline\hline
\end{tabular}
\caption{The pseudoscalar- and vector-meson trajectories in the strange
sector~\cite{Brisudova:1999ut,Titov:2008yf}.} 
\label{TAB:1}
\end{table}

With the Regge trajectories determined, the energy-scale parameters
$s_K^{K N : \phi \Lambda}$ and $s_{K^*}^{K N : \phi \Lambda}$ in Eq.~(\ref{eq:RegAmpl1}) can
be derived using the corresponding scale parameters for the diagonal transitions
$K N \to K N (s^{K N})$ and $\phi \Lambda \to \phi \Lambda (s^{\phi \Lambda})$
as~\cite{Boreskov:1983bu,Kaidalov:1986zs,Kaidalov:1994mda}
\begin{align}
(s_K^{K N : \phi \Lambda})^{2(\alpha_K(0))}
= &\, (s^{K N})^{\alpha_\pi(0)} \times (s^{\phi \Lambda})^{\alpha_{\eta_s}(0)}
\cr
(s_{K^*}^{K N : \phi \Lambda})^{2(\alpha_{K^*}(0)-1)}
= &\, (s^{K N})^{\alpha_\rho(0)-1} \times (s^{\phi \Lambda})^{\alpha_\phi(0)-1},
\label{eq:EneScaPara1}
\end{align}
where $s^{ab}$ is proportional to the total transverse masses of the constituent
quarks in hadrons $a$ and $b$,
\begin{align}
s^{ab} = \left( \sum_i m_{\perp i} \right)_a \left( \sum_j m_{\perp j} \right)_b.
\end{align}
We use $m_{\perp q} \simeq 0.5$ GeV, $m_{\perp s} \simeq 0.6$ GeV, and $m_{\perp c}
\simeq 1.6$ GeV, which lead to
\begin{align}
& s^{K N} = 1.65, \,\,\, s^{\phi \Lambda} = 1.92,
\cr
& s_K^{K N : \phi \Lambda} = 1.91, \,\,\,
s_{K^*}^{K N : \phi \Lambda} = 1.82,
\end{align}
in units of ${\rm GeV^2}$.

\subsection{Charm production: $K^- p \to D_s^{*-} \Lambda_c^+,\,J/\psi \Lambda$}
\label{Sec:II-2}

We extend the analysis of strangeness production in $K^- p \to \phi \Lambda$ to
charm production by simply replacing strange hadrons with charm hadrons.

\subsubsection{$K^- p \to D_s^{*-} \Lambda_c^+$}

The $t$-channel Regge amplitudes for the $K^- p \to D_s^{*-} \Lambda_c^+$ reaction are
obtained by replacing $K \to D (1869,0^-)$ and $K^* \to D^* (2007,1^-)$ in the
propagators, and $\phi \to D_s^* (2106,1^-)$ and $\Lambda \to \Lambda_c (2286,1/2^+)$
in the final states (Fig~\ref{FIG01}(a)):
\begin{align}
T_D (s,t) = &\, \mathcal M_D(s,t)
\left( \frac{s}{s_D^{K N : D_s^* \Lambda_c}} \right)^{\alpha_D(t)}
\cr
& \times \Gamma (-\alpha_D(t)) \alpha_D' F_{PS}^2(t),
\cr
T_{D^*} (s,t) = &\, \mathcal M_{D^*}(s,t)
\left( \frac{s}{s_{D^*}^{K N : D_s^* \Lambda_c}} \right)^{\alpha_{D^*}(t)-1}
\cr
& \times \Gamma (1-\alpha_{D^*}(t)) \alpha_{D^*}' F_V^2(t).
\label{eq:RegAmpl2}
\end{align}
The $D$- and $D^*$-Regge trajectories are determined as in the strangeness
production case.
The relevant values are listed in Table~\ref{TAB:2}~\cite{Brisudova:1999ut}.
\begin{table}[h]
\begin{tabular}{cccc}
\hline\hline
&$\alpha(0)$
&\hspace{1.1em}$\sqrt{T}\,[{\rm GeV}]$\hspace{1.1em}
&$\alpha'\,[{\rm GeV}^{-2}]$
\\\hline
$\bar{q}q(\pi)$    \hspace{6.0em}& -0.0118  & 2.82 & 0.647   \\
$\bar{c}q(D)$      \hspace{6.0em}& -1.61105 & 4.16 & 0.439   \\
$\bar{c}c(\eta_c)$ \hspace{6.0em}& -3.2103  & 5.49 & 0.332   \\
\hline
$\bar{q}q(\rho)$   \hspace{6.0em}&  0.55 & 2.46 & 0.742   \\
$\bar{c}q(D^*)$    \hspace{6.0em}& -1.02 & 3.91 & 0.467   \\
$\bar{c}c(J/\psi)$ \hspace{6.0em}& -2.60 &5.36  & 0.340   \\
\hline\hline
\end{tabular}
\caption{Pseudoscalar- and vector-meson trajectories in the charm
sector~\cite{Brisudova:1999ut}.}
\label{TAB:2}
\end{table}

Equation~(\ref{eq:EneScaPara1}) can be also modified as~\cite{Boreskov:1983bu,
Kaidalov:1986zs,Kaidalov:1994mda}
\begin{align}
(s_D^{K N : D_s^* \Lambda_c})^{2 \alpha_D(0)}
= &\, (s^{K N})^{\alpha_\pi(0)} \times (s^{D_s^* \Lambda_c})^{\alpha_{\eta_c} (0)}
\cr
(s_{D^*}^{K N : D_s^* \Lambda_c})^{2(\alpha_{D^*}(0)-1)}
= &\, (s^{K N})^{\alpha_\rho(0)-1} \times (s^{D_s^* \Lambda_c})^{\alpha_{J/\psi}(0)-1},
\label{eq:EneScaPara2}
\end{align}
where
\begin{align}
& s^{K N} = 1.65, \,\,\, s^{D_s^* \Lambda_c} = 5.72,
\cr
& s_D^{K N : D_s^* \Lambda_c} = 5.69, \,\,\,
s_{D^*}^{K N : D_s^* \Lambda_c} = 5.00,
\end{align}
in units of ${\rm GeV^2}$.

The coupling constants in Eq.~(\ref{eq:Ampl1}) are replaced as $g_{\phi K K} \to
g_{D_s^* D K}$ and $g_{\phi K^* K} \to g_{D_s^* D^* K}$.
The coupling $g_{D_s^* D K}$ is fixed using SU(4) flavor symmetry as $g_{D_s^* D K} =
g_{\rho \pi \pi} / \sqrt{2}$~\cite{Haidenbauer:2014rva}, where $g_{\rho \pi \pi} = 5.94$.
The strength of the $D_s^* D^* K$ vertex is then estimated from heavy-quark spin
symmetry, which relates it to $g_{D_s^* D K}$ through $g_{D_s^* D^* K} = - g_{D_s^* D K} /
\sqrt{M_{D} M_{D^*}} = - 1.53\, \rm{GeV^{-1}}$~\cite{Colangelo:2003sa,Guo:2010ak}.
For the meson--baryon couplings, we assume SU(4) symmetry and take them to be
identical to those in the strange sector, namely $g_{D N \Lambda_c} = g_{K N \Lambda}$,
$g_{D^* N \Lambda_c} = g_{K^* N \Lambda}$, and $\kappa_{D^* N \Lambda_c} = \kappa_{K^* N \Lambda}$.
In the tensor term, the same mass scale $M_N$ as in the strange sector is employed.

\subsubsection{$K^- p \to J/\psi \Lambda$}

The $t$-channel Regge amplitudes for the $K^- p \to J/\psi \Lambda$ reaction are
obtained just by replacing $\phi \to J/\psi (3096, 1^-)$ in the final state
(Fig~\ref{FIG01}(b)):
\begin{align}
T_K  (s,t) = &\, \mathcal M_K(s,t)
\left( \frac{s}{s_K^{K N : J/\psi \Lambda}} \right)^{\alpha_K(t)}
\cr
& \times \Gamma (-\alpha_K(t)) \alpha_K' F_{PS}^2(t),
\cr
T_{K^*} (s,t) = &\, \mathcal M_{K^*}(s,t)
\left( \frac{s}{s_{K^*}^{K N : J/\psi \Lambda}} \right)^{\alpha_{K^*}(t)-1}
\cr
& \times \Gamma (1-\alpha_{K^*}(t)) \alpha_{K^*}' F_V^2(t).
\label{eq:RegAmpl3}
\end{align}
The energy scale parameters are assumed to be the same as those in the strange
sector:
$s_K^{K N : J/\psi \Lambda} = s_K^{K N : \phi \Lambda}$ and
$s_{K^*}^{K N : J/\psi \Lambda} = s_{K^*}^{K N : \phi \Lambda}$.

The coupling constants in Eq.~(\ref{eq:Ampl1}) are replaced as
$g_{\phi K K} \to g_{J/\psi K K}$ and $g_{\phi K^* K} \to g_{J/\psi K^* K}$.
Since the branching ratios of $J/\psi$ to the $K$ and $K^*$ are known
experimentally as
\begin{align}
{\mathcal B} (J/\psi \to K^+ K^-) =&\, 0.0306\,\%,
\cr
{\mathcal B} (J/\psi \to K^+ K^{*-}) =&\, 0.60\,\%,
\end{align}
we can get their coupling constants as
\begin{align}
g_{J/\psi K K} =&\, 1.28 \cdot 10^{-3},
\cr
g_{J/\psi K^* K} =&\, 2.85 \cdot 10^{-3} \, {\rm GeV^{-1}},
\end{align}
using the width of $\Gamma_{J/\psi} = 92.6$ keV and the relations,
\begin{align}
\Gamma (J/\psi \to K^+ K^-) =&\, \frac{q_K^3}{6 \pi M_{J/\psi}^2} g_{J/\psi K K}^2,
\cr
\Gamma (J/\psi \to K^+ K^{*-}) =&\, \frac{q_{K^*}^3}{12 \pi} g_{J/\psi K^* K}^2,
\label{eq:DWidth}
\end{align}
where $q_{K(K^*)}$ is the magnitude of the three-momentum of $K(K^*)$ in the rest
frame of $J/\psi$.

\subsection{$s$- and $u$-channel contributions to $K^- p \to \phi \Lambda$}
\label{Sec:II-3}

\begin{figure}[h]
\centering
\includegraphics[scale=0.55]{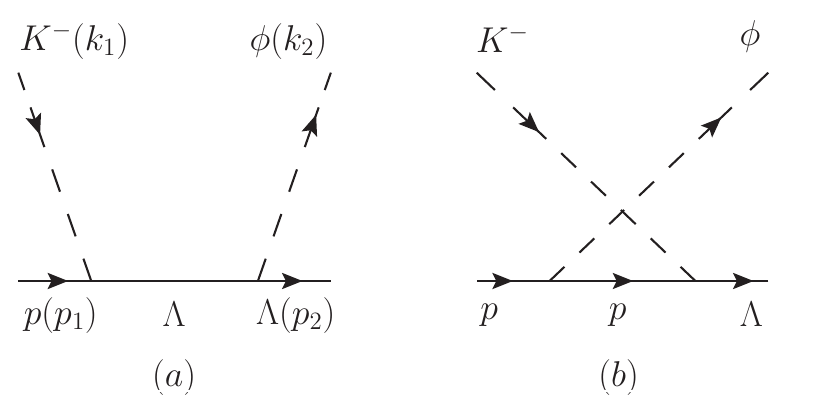}
\caption{Feynman diagrams for $K^- p \to \phi \Lambda$: (a) $\Lambda$ exchange in the
$s$ channel and (b) nucleon exchange in the $u$ channel.}
\label{FIG03}
\end{figure}

The $s$-channel $\Lambda$ [Fig.~\ref{FIG03}(a)] and $u$-channel nucleon
[Fig.~\ref{FIG03}(b)] exchanges may contribute to the $K^- p \to \phi \Lambda$
reaction, and they are examined in this work.

The effective Lagrangians for the $\phi$-meson–baryon interaction vertices are written
as
\begin{align}
\mathcal{L}_{\phi \Lambda \Lambda} =&\, - g_{\phi \Lambda \Lambda} \bar \Lambda
\left[ \gamma_\mu \Lambda - \frac{\kappa_{\phi \Lambda \Lambda}}{2M_N} \sigma_{\mu\nu}
\Lambda \partial^\nu \right] \phi^\mu,
\cr
\mathcal{L}_{\phi N N} =&\, - g_{\phi N N} \bar N
\left[ \gamma_\mu N - \frac{\kappa_{\phi N N}}{2M_N} \sigma_{\mu\nu}
N \partial^\nu \right] \phi^\mu,
\label{eq:Lag3}
\end{align}
where the Nijmegen soft-core model (NSC97a)~\cite{Rijken:1998yy,Stoks:1999bz} is used
to determine the coupling constants,
\begin{align}
g_{\phi \Lambda \Lambda} = &\, -3.80, \,\,\, \kappa_{\phi \Lambda \Lambda} = 1.78,
\cr
g_{\phi N N} = &\, -1.47, \,\,\, \kappa_{\phi N N} = -2.64.
\label{eq:Coupl2}
\end{align}
The scattering amplitudes corresponding to the $s$-channel $\Lambda$ and $u$-channel
nucleon exchanges are, respectively, given by
\begin{align}
\mathcal M_\Lambda^\mu =&\,
i \frac{g_{\phi \Lambda \Lambda}}{s-M_\Lambda^2} \frac{g_{K N \Lambda}}{M_N + M_\Lambda}
\left [ \gamma^\mu - \frac{i\kappa_{\phi \Lambda \Lambda}}{2 M_N} \sigma^{\mu\nu} k_{2 \nu}
\right ]
\cr & \times
(\rlap{/}{k_1}+\rlap{/}{p_1}+M_\Lambda) \gamma^\alpha \gamma_5 k_{1\alpha},
\cr
\mathcal M_N^\mu =&\,
i \frac{g_{\phi N N}}{u-M_N^2}
\frac{g_{K N \Lambda }}{M_N + M_\Lambda} \gamma^\alpha \gamma_5
(\rlap{/}{p_2}-\rlap{/}{k_1}+M_N)
\cr & \times
\left [ \gamma^\mu - \frac{i\kappa_{\phi N N}}{2 M_N} \sigma^{\mu\nu} k_{2 \nu}
\right ] k_{1\alpha}.
\label{eq:Ampl2}
\end{align}
with $\mathcal M = \varepsilon_\mu^* \bar{u}_\Lambda \, \mathcal M^\mu \,u_N$.
The relevant hadrons are spatially extended, so we consiter the following form factor
for each vertex:
\begin{align}
F_\Lambda (s) =&\, \frac{\Lambda_s^4}{\Lambda_s^4 + (s - M_\Lambda^2)^2},
\cr
F_N (u) =&\, \frac{\Lambda_u^4}{\Lambda_u^4 + (u - M_N^2)^2},
\label{eq:FormFac_su}
\end{align}
for the $s$-channel and $u$-channel diagrams, respectively.
The cutoff masses are determined to be $\Lambda_s = \Lambda_u = 0.8$ GeV.

\subsection{Pentaquark contributions to $K^- p \to
D_s^{*-} \Lambda_c^+,\,J/\psi \Lambda$}
\label{Sec:II-4}

\begin{figure}[h]
\centering
\includegraphics[scale=0.60]{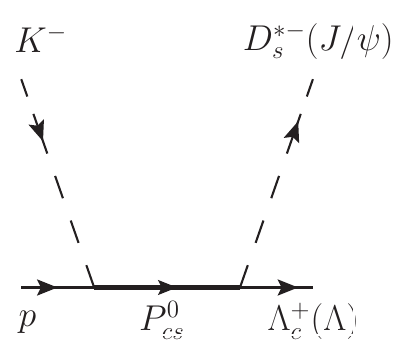}
\caption{Feynman diagram for the $K^- p \to D_s^{*-} \Lambda_c^+$, $J/\psi \Lambda$
reactions via $s$-channel $P_{cs}^0$ pentaquark exchange.}
\label{FIG04}
\end{figure}

We now examine the $s$-channel contributions from the hidden-charm pentaquark states
with strangeness, $P_{cs}(4337)^0$~\cite{LHCb:2022ogu} and
$P_{cs}(4459)^0$~\cite{LHCb:2020jpq}, as shown in Fig.~\ref{FIG04}.
We point out that $P_{cs}(4337)^0$ exchange is allowed only for $K^- p \to J/\psi
\Lambda$, because the $D_s^{*-}\Lambda_c^+$ production threshold lies above the
$P_{cs}(4337)^0$ mass.

While the spin-parity of $P_{cs}(4337)^0$ has been determined to be $J^P=1/2^-$, that
of $P_{cs}(4459)^0$ is still unsettled.
In a hadronic molecular interpretation, Ref.~\cite{Peng:2020hql} favors $J^P = 3/2^-$
over $1/2^-$, which is also supported by Ref.~\cite{Chen:2021tip}.
In contrast, within a unitarized approach, $J^P=1/2^-$ is preferred over
$3/2^-$~\cite{Lu:2021irg}.
Accordingly, we consider both $J^P = 1/2^-$ and $3/2^-$ assignments for
$P_{cs}(4459)^0$ in our analysis.
The branching ratios are taken as follows~\cite{Xiao:2021rgp}:
\begin{align}
{\mathcal B} (P_{cs} (4459,\,1/2^-) \to \bar D_s^* \Lambda_c) =&\, 80.86\,\%,
\cr
{\mathcal B} (P_{cs} (4459,\,3/2^-) \to \bar D_s^* \Lambda_c) =&\, 81.64\,\%,
\cr
{\mathcal B} (P_{cs} (4459,\,1/2^-) \to J/\psi \Lambda) =&\, 3.31\,\%,
\cr
{\mathcal B} (P_{cs} (4459,\,3/2^-) \to J/\psi \Lambda) =&\, 14.68\,\%.
\label{eq:BR_Pcs_1}
\end{align}
For $P_{cs}(4337)^0$, we adopt the branching ratio quoted in
Ref.~\cite{Wang:2024rsm},
\begin{align}
{\mathcal B} (P_{cs} (4337) \to J/\psi \Lambda) = 84.7\,\%.
\label{eq:BR_Pcs_2}
\end{align}
Note that since no information is currently available for the $P_{cs}\to \bar K N$
branching ratios of $P_{cs}(4337)^0$ and $P_{cs}(4459)^0$, we treat them as free
parameters in the present analysis.

The effective Lagrangians for the $P_{cs} N K$ vertex are given
by~\cite{Kim:2024mqx}
\begin{align}
\mathcal L_{PNK}^{1/2^\pm} &=
-i g_{PNK} \bar N \Gamma^{(\pm)} P K + {\rm H.c.},
\cr
\mathcal L_{PNK}^{3/2^\pm} &=
\frac{g_{PNK}}{M_K} \bar N \Gamma^{(\mp)} P^\mu \partial_\mu K + {\rm H.c.},
\label{eq:ResLag1}
\end{align}
where the off-shell part of the Rarita-Schwinger fields is also ignored because the
resonances are almost on mass shell.
For the $K^- p \to D_s^{*-} \Lambda_c^+$ reaction, the effective Lagrangians for the
$P_{cs} \Lambda_c D_s^*$ vertex can be expressed
as~\cite{Kim:2024mqx}
\begin{align}
\mathcal{L}_{P \Lambda_c D_s^*}^{1/2^\pm}
&= - g_{P \Lambda_c D_s^*} \bar \Lambda_c \Gamma_\mu^{(\mp)} P D_s^{*\mu} + {\rm H.c.},
\cr
\mathcal{L}_{P \Lambda_c D_s^*}^{3/2^\pm}
&= -\frac{i g_{P \Lambda_c D_s^*}}{2M_N}
\bar \Lambda_c \Gamma_\nu^{(\pm)} P_\mu D_s^{*\mu\nu} + {\rm H.c.}, 
\label{eq:ResLag2}
\end{align}
where $D_s^{*\mu\nu} = \partial^\mu D_s^{*\nu} - \partial^\nu D_s^{*\mu}$.
The heavy pentaquark states play a dominant role near the threshold region, so
additional terms are ignored.
$P$ denotes the field of the pentaquark state $P_{cs}$.
For the $K^- p \to J/\psi \Lambda$ reaction, we can just replace the hadrons as
$D_s^* \to J/\psi$ and $\Lambda_c \to \Lambda$.

The following notations are used:
\begin{align}
\Gamma^{(\pm)} = \left(
\begin{array}{c} 
\gamma_5 \\ \mathbf{1}
\end{array} \right),
\qquad
\Gamma_\mu^{(\pm)} = \left(
\begin{array}{c}
\gamma_\mu \gamma_5 \\ \gamma_\mu 
\end{array} \right).
\end{align}

The scattering amplitudes for the exchanges of the pentaquark states are written
as
\begin{align}
\mathcal{M}_{P(1/2^+)}^\mu
=&\, i \frac{g_{PNK}\, g_{P \Lambda_c D_s^*}}{s-M_P^2} \gamma^\mu
(\slashed{q}_s + M_P) \gamma_5,
\cr
\mathcal{M}_{P(1/2^-)}^\mu
=&\, i \frac{g_{PNK}\, g_{P \Lambda_c D_s^*}}{s-M_P^2} \gamma^\mu \gamma_5
(\slashed{q}_s + M_P),
\cr
\mathcal{M}_{P(3/2^+)}^\mu
=&\, i\frac{g_{PNK}}{M_K} \frac{g_{P \Lambda_c D_s^*}}{2M_N}
\frac{1}{s-M_P^2} \gamma_5 \gamma_\nu
\cr     &
\times (k_2^\alpha g^{\mu\nu} - k_2^\nu g^{\mu\alpha})
\Delta_{\alpha\beta}(q_s, M_P) k_1^\beta,
\cr
\mathcal{M}_{P(3/2^-)}^\mu
=&\, i\frac{g_{PNK}}{M_K} \frac{g_{P \Lambda_c D_s^*}}{2M_N}
\frac{1}{s-M_P^2} \gamma_\nu
\cr     &
\times (k_2^\alpha g^{\mu\nu} - k_2^\nu g^{\mu\alpha})
\Delta_{\alpha\beta}(q_s, M_P) k_1^\beta \gamma_5,
\cr
\label{eq:InvAmp2}
\end{align}
with $\mathcal M = \varepsilon_\mu^* \bar{u}_\Lambda \, \mathcal M^\mu \,u_N$ and
$q_s = k_1 + p_1$.
The spin-3/2 projection operator is given by
\begin{align}
\Delta_{\alpha\beta}(p, M) =&\, (\slashed{p} + M)
\biggl[ -g_{\alpha\beta} + \frac13 \gamma_\alpha \gamma_\beta
\cr
& + \frac{1}{3M} (\gamma_\alpha p_\beta - p_\alpha \gamma_\beta) + \frac{2}{3M^2} p_\alpha
p_\beta \biggr].
\label{eq:FF}
\end{align}
Given the decay widths of the $P_{cs}$ states, the propagators of the pentaquark
states should be modified to $M_{P_{cs}} \to (M_{P_{cs}} - i\Gamma_{P_{cs}}/2$).

From the values of branching ratios given in Eqs.~(\ref{eq:BR_Pcs_1}) and
(\ref{eq:BR_Pcs_2}), the coupling constants for the $P_{cs}$ interactions are
obtained and are summarized in Table~\ref{TAB:3}~\cite{Kim:2024mqx}.

\begin{table}[h]
\begin{tabular}{ccc}
\hline\hline
&\hspace{0.8em}$g_{P \Lambda_c D_s^*}$\hspace{0.8em}
&\hspace{0.8em}$g_{P \Lambda J/\psi}$\hspace{0.8em}
\\\hline
$P_{cs} (4459,\,1/2^-)$ & 0.382 & 0.0802     \\
$P_{cs} (4459,\,3/2^-)$ & 0.589 & 0.176     \\
$P_{cs} (4380)$         & $-$     & 0.313     \\
\hline\hline
\end{tabular}
\caption{Coupling constants of $P_{cs}(4459)$ to $D_s^* \Lambda_c$ and $J/\psi
\Lambda$ for each $J^P = 1/2^-$ and $3/2^-$ assignment, together with that of
$P_{cs}(4380)$ to $J/\psi \Lambda$.}
\label{TAB:3}
\end{table}

We consider the form factor 
\begin{align}
F_{P_{cs}}(s) = \frac{\Lambda_P^4}{\Lambda_P^4 + (s-M_{P_{cs}}^2)^2},
\label{eq:FF}
\end{align}
at each vertex, where the cutoff masses are selected as $\Lambda_P = $ 1.0 GeV.
This cutoff mass does not play a crucial role in the present calculation, since
the $P_{cs}$ states lie close to the reaction threshold.

\subsection{Spin density matrix elements}
\label{Sec:II-5}

The SDMEs, which characterize the polarization states of the relevant hadrons,
provide key observables for understanding the reaction mechanism~\cite{Kim:2017hhm}.
In this work, we focus on the case in which only the produced vector meson is
polarized.
For definiteness, we consider the decay channel $\phi \to K^+ K^-$ in the $K^- p \to
\phi \Lambda$ reaction.
The analysis of the outgoing $K^+$ in the vector-meson rest frame involves an
ambiguity in the choice of the quantization axis.
One possible choice is to take the axis antiparallel to the momentum
of the outgoing hyperon $\Lambda$ in the $\phi$ decay.
Alternatively, it may be defined to be parallel to the momentum of the incoming
meson, i.e., along the initial beam direction.

Following the convention of Refs.~\cite{Crennell:1972km,Schilling:1969um}, the
former choice is referred to as the helicity (H) frame, while the latter is known
as the Gottfried–Jackson (GJ) frame.
The helicity frame is commonly used to test $s$-channel helicity conservation,
whereas the GJ frame is suitable for investigating the $t$-channel exchange
mechanism.

The decay probabilities are expressed in terms of the spin density matrix elements
$\rho_{\lambda \lambda'}$, where the vector-meson helicity $\lambda_V$ is abbreviated as
$\lambda$.
These SDMEs are determined by the reaction amplitudes given by
Eq.~(\ref{eq:RegAmpl1})
\begin{align}
\rho_{\lambda\lambda'} = \frac{1}{{\mathcal N}^2}
\sum\limits_{s_f = \pm\frac12,\, s_i = \pm\frac12}
T_{\lambda,s_f;s_i}\, T^*_{\lambda',s_f;s_i},
\label{eq:SDME}
\end{align}
with the normalization factor
\begin{align}
{\mathcal N}^2 = \sum_{\lambda_,s_f,s_i} |T_{\lambda,s_f ; s_i}|^2.
\label{eq:SDME-NF}
\end{align}
We make use of the Hermitian conditions,
\begin{align}
\rho_{1-1} = \rho_{-11}, \,\,\,
\rho_{10} = \rho_{01}, \,\,\,
\rho_{-10} = \rho_{0-1}.
\label{eq:SDME}
\end{align}
In addition, the normalization condition $\rho_{00}+\rho_{11}+\rho_{-1-1}=1$ and the
symmetry conditions,
\begin{align}
&\rho_{11} = \rho_{-1-1},\,\,\, \rho_{\pm 10} = \rho_{0 \pm 1},
\cr
&\rho_{1-1} = \rho_{-11},\,\,\, \rho_{\pm 10} = -\rho_{0 \mp 1},
\label{eq:SDME}
\end{align}
are satisfied in our numerical calculations.

The decay angular distributions can be expressed in terms of the SDMEs
as
\begin{align}
W^{0}(\Omega_f) =&\, \frac{3}{4\pi} \Bigl[ \rho_{00}\cos^2\Theta +
\rho_{11}\sin^2\Theta
- \rho_{1-1} \sin^2\Theta \cos2\Phi
\cr &
- {\sqrt{2}}\, {\rm Re}(\rho_{10}) \sin2\Theta\cos\Phi \Bigr],
\label{eq:DAD}
\end{align}
denoting the polar and the azimuthal angles of the outgoing pseudoscalar $K^+$ meson
by $\Theta$ and $\Phi$, respectively~\cite{Kim:2017hhm}.

\section{Numerical Results}
\label{Sec:III}

\subsection{Strangeness production: $K^- p \to \phi \Lambda$}
\label{Sec:III-1}

We first present our numerical results for the $K^- p \to \phi \Lambda$ reaction,
considering the $K$- and $K^*$-Reggeon exchanges discussed in Sec.~\ref{Sec:II-1}.
The cutoff masses in the form factor of Eq.~(\ref{eq:FormFac_t}) are fixed by
reproducing the available data,
\begin{align}
\Lambda_{PS} = 0.5\,{\rm GeV},\,\,\, \Lambda_V = 0.8\,{\rm GeV}.
\label{eq:SDME}
\end{align}

\begin{figure}[ht]
\centering
\includegraphics[width=7.0cm]{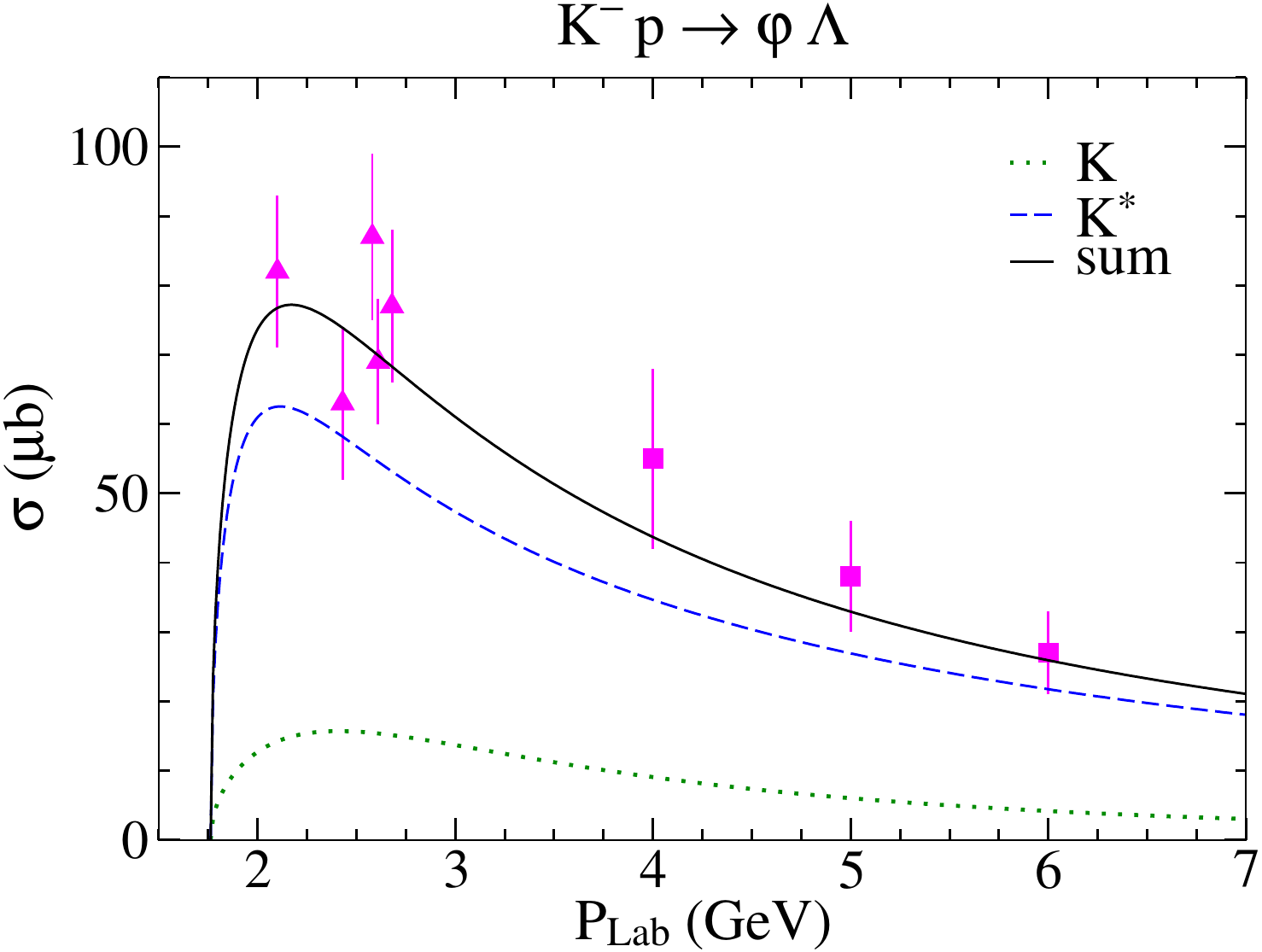}
\caption{Total cross section for the $K^- p \to \phi \Lambda$ reaction as a function
of $P_{\rm Lab}$.
The green dotted and blue dashed curves denote the contributions from $K$- and
$K^*$-Reggeon exchanges, respectively, while the black solid curve represents their
sum.
Experimental data are taken from Ref.~\cite{Lindsey:1966zz} (triangles) and
\cite{Ayres:1974aj} (squares).}
\label{FIG05}
\end{figure}
Figure~\ref{FIG05} displays the total cross section as a function of the laboratory
beam energy, $P_{\rm Lab}$.
The $K^*$-Reggeon exchange clearly dominates over the $K$-Reggeon exchange, and its
contribution becomes increasingly important at higher beam energies.
The coherent sum of the two exchanges provides a good overall description of the
experimental data~\cite{Lindsey:1966zz,Ayres:1974aj}.

\begin{figure}[ht]
\centering
\includegraphics[width=\columnwidth]{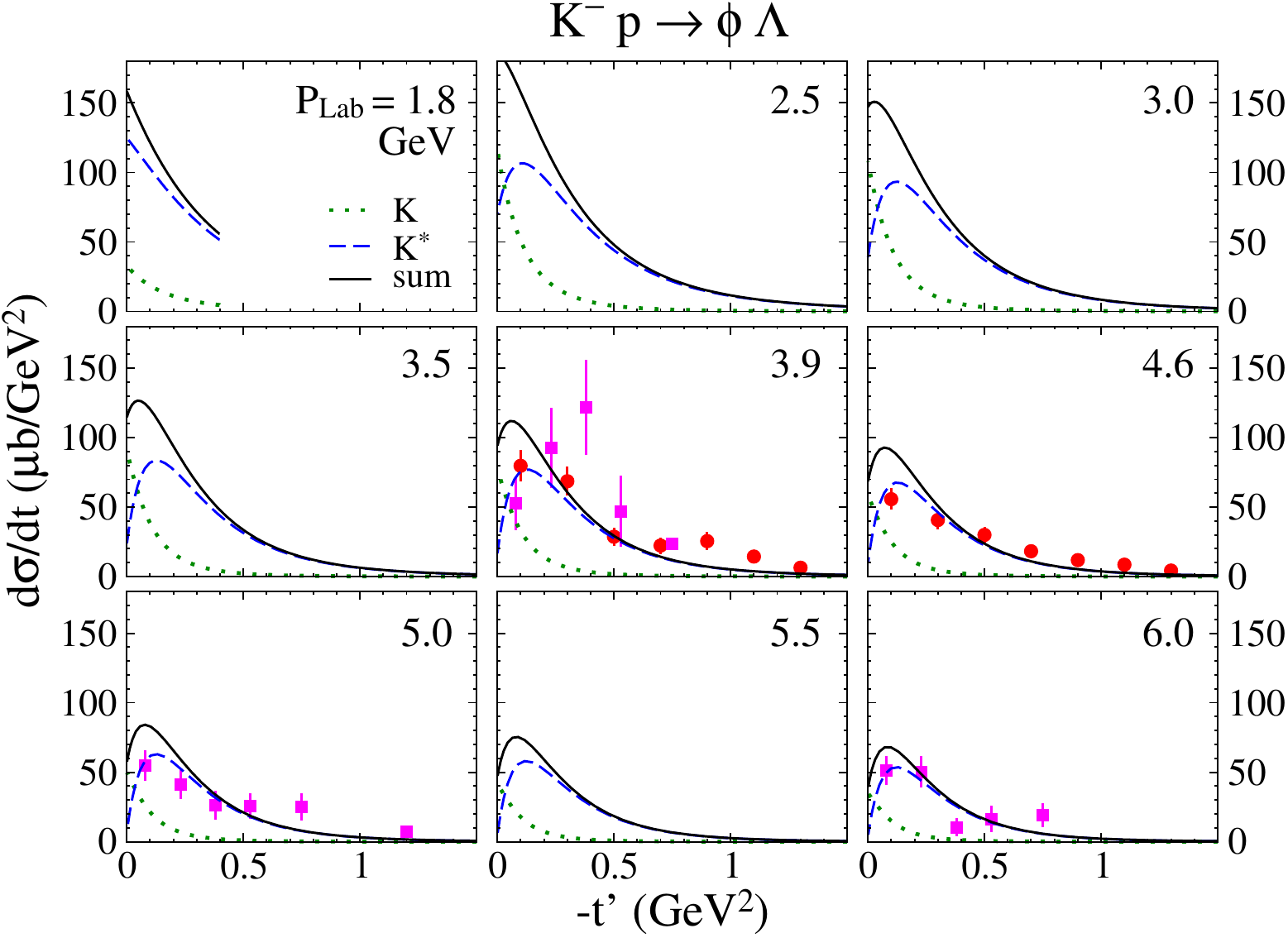}
\caption{$t$-dependent differential cross sections for the $K^- p \to \phi \Lambda$
reaction at nine fixed beam energies.
Experimental data are taken from Refs.~\cite{Aguilar-Benitez:1972ngz} (circles)
and \cite{Ayres:1974aj} (squares).
Curve notations are the same as in Fig.~\ref{FIG05}.}
\label{FIG06}
\end{figure}
We present the differential cross sections as functions of $-t' = -t + t_{min}$ in
Fig.~\ref{FIG06} at nine fixed beam energies.
As discussed in Eq.~(\ref{eq:Asym:Ampl}), the two Reggeon exchanges exhibit
qualitatively different behaviors at very forward angles.
The available experimental data strongly favor the $K^*$-Reggeon exchange over the 
$K$-Reggeon exchange.
A larger $K$-Reggeon contribution would worsen the agreement~\cite{
Aguilar-Benitez:1972ngz,Ayres:1974aj}.

\begin{figure}[ht]
\centering
\includegraphics[width=4.0cm]{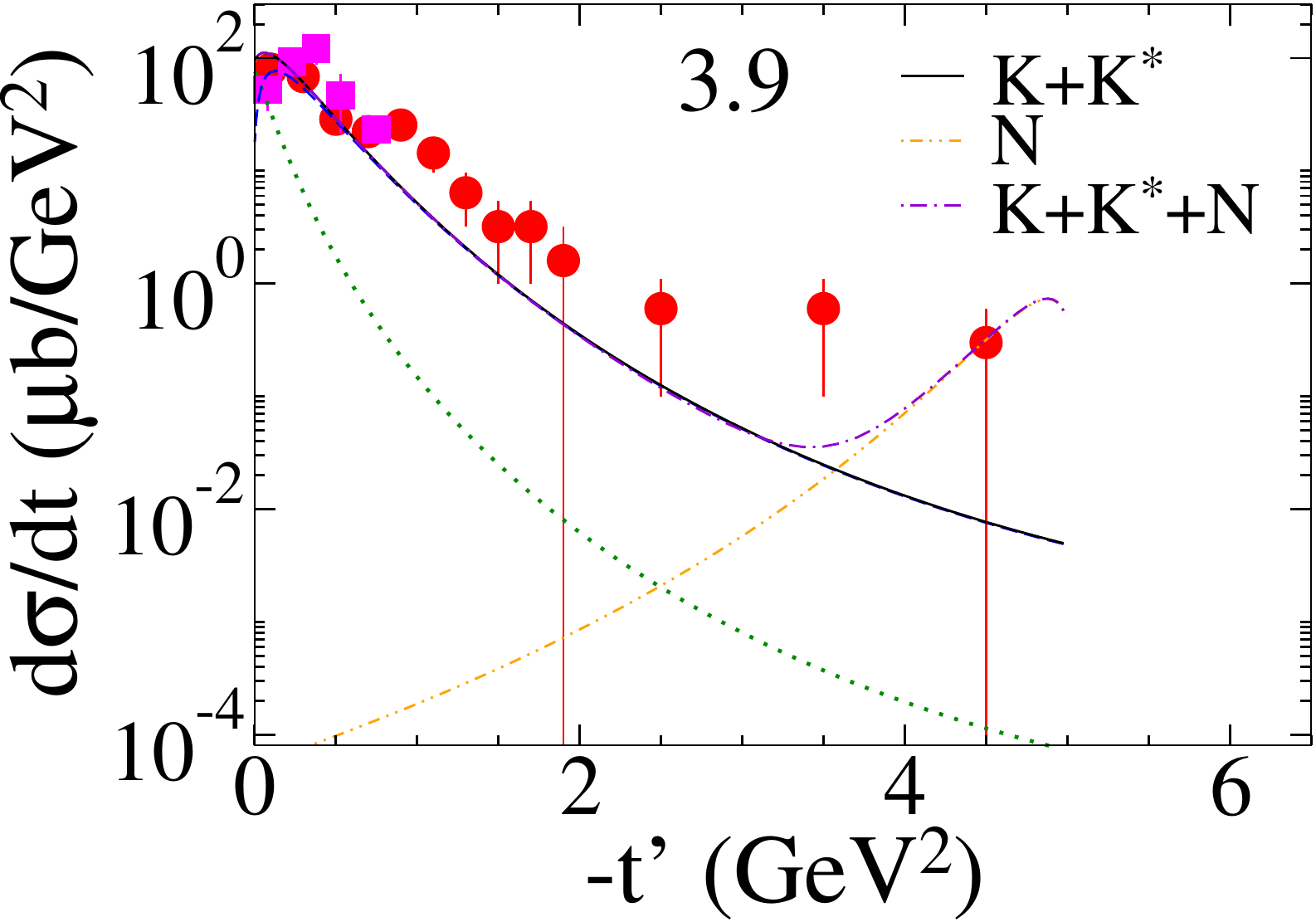}\,
\includegraphics[width=4.0cm]{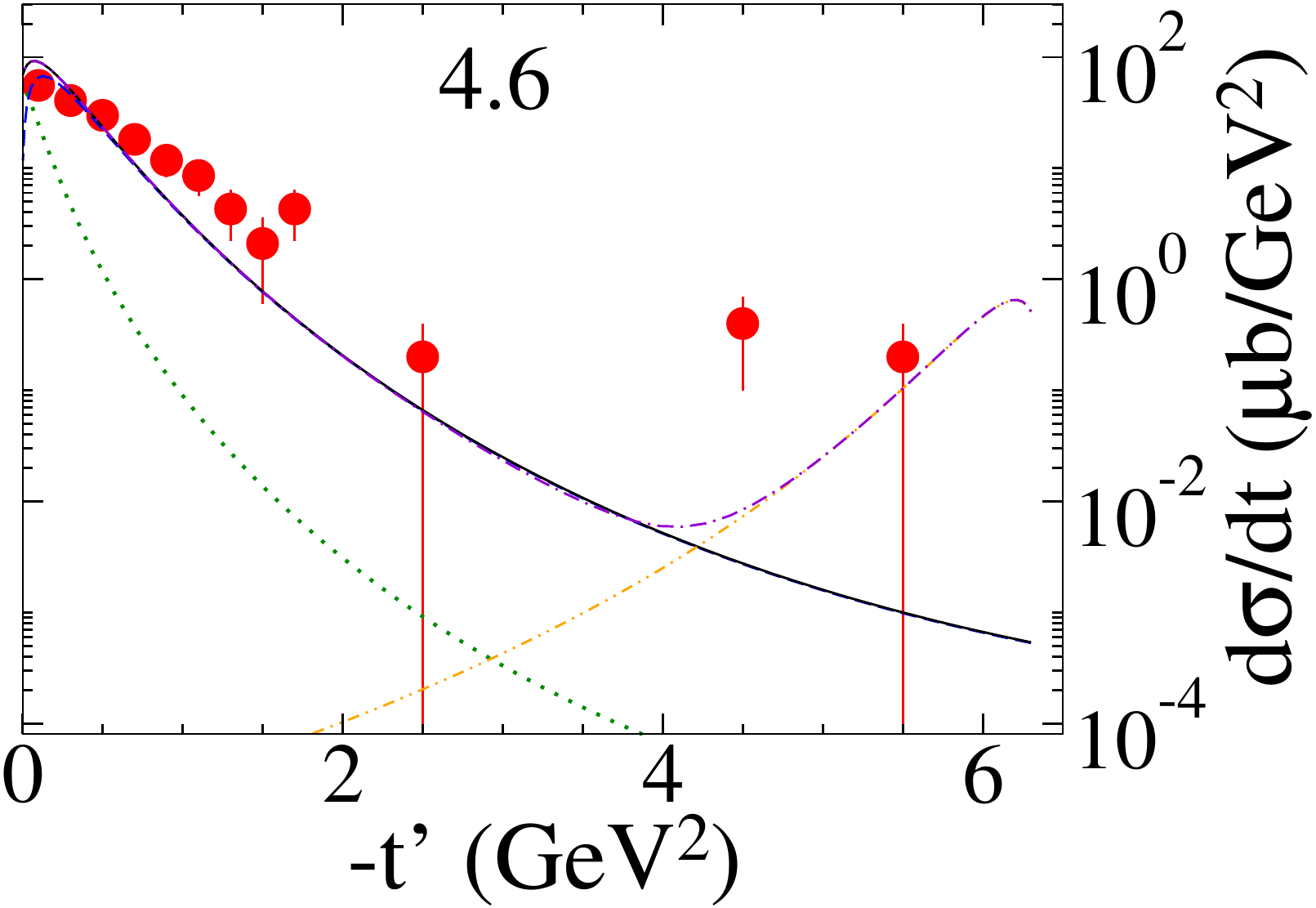}
\caption{$t$-dependent differential cross sections for the $K^- p \to \phi \Lambda$
reaction at $P_{\rm Lab}=3.9$ and $4.6$ GeV, shown on a logarithmic scale.
Experimental data are taken from Refs.~\cite{Aguilar-Benitez:1972ngz} (circles) and
\cite{Ayres:1974aj} (squares).
In addition to $t$-channel $K$- and $K^*$-Reggeon exchanges, $u$-channel nucleon
exchange is also included.}
\label{FIG07}
\end{figure}
In Fig.~\ref{FIG07}, we examine additional contributions from the $s$-channel
$\Lambda$ and $u$-channel nucleon exchanges, taking advantage of the backward-angle
scattering data reported in Ref.~\cite{Aguilar-Benitez:1972ngz}.
The $K^*$-Reggeon exchange reproduces the slope of the experimental data well, which
supports the validity of our Regge formalism.
Owing to the different intercepts of the Regge trajectories, $\alpha_K(0) <
\alpha_{K^*} (0)$, the $K$-Reggeon exchange decreases more rapidly with increasing
$-t'$ than the $K^*$-Reggeon exchange, as indicated in Eq.~(\ref{eq:Asym:dsdt}).
The small deviations observed at $-t' \geqslant 4 \,\rm{GeV}^2$ can be accounted for
by including the $u$-channel nucleon contribution.
In contrast, the $s$-channel $\Lambda$ exchange interferes destructively with the 
$t$-channel contribution at backward angles and is therefore strongly suppressed in
this mechanism.

\begin{figure*}[hb]
\centering
\includegraphics[width=8.5cm]{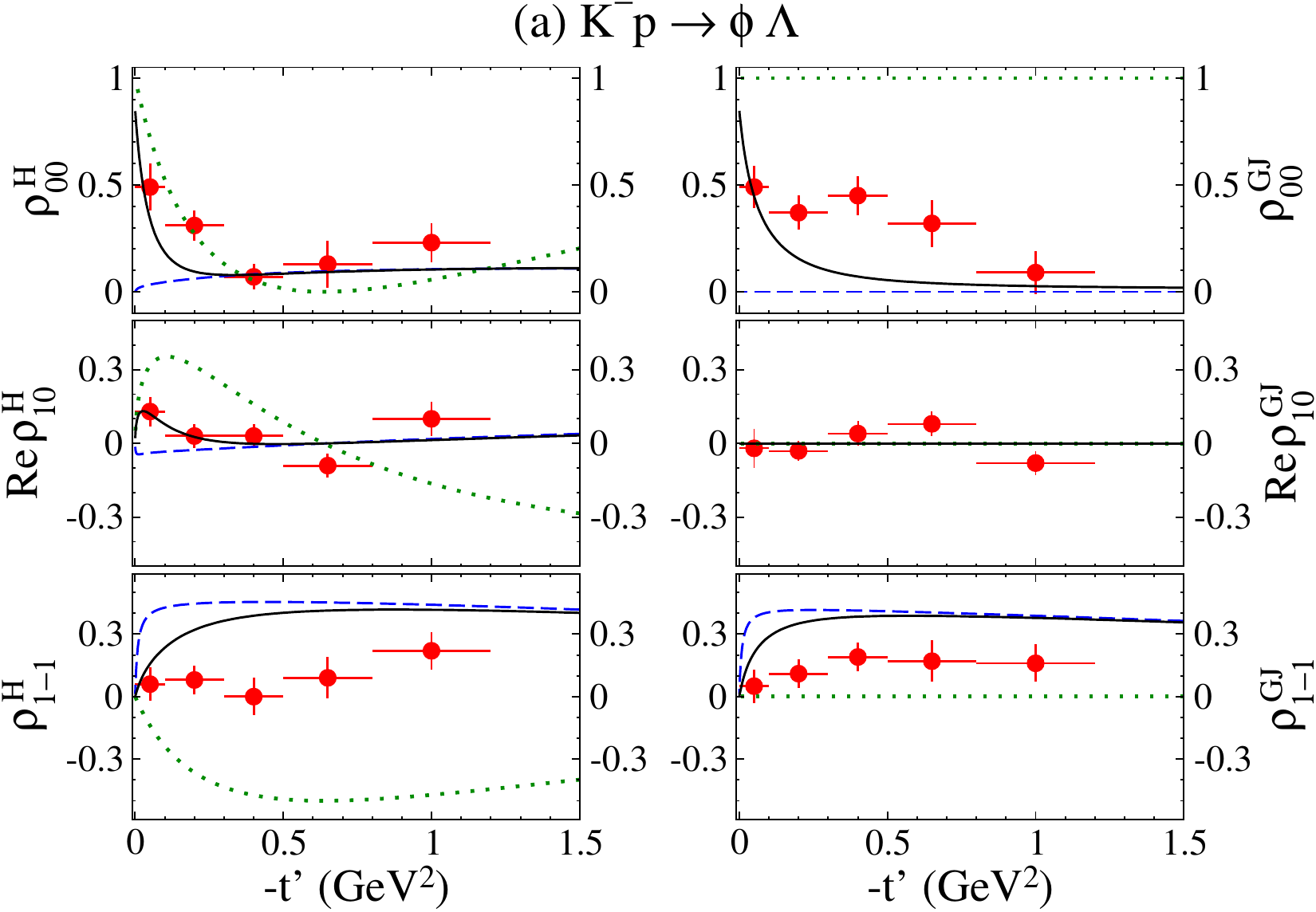} \hspace{1.5em}
\includegraphics[width=8.5cm]{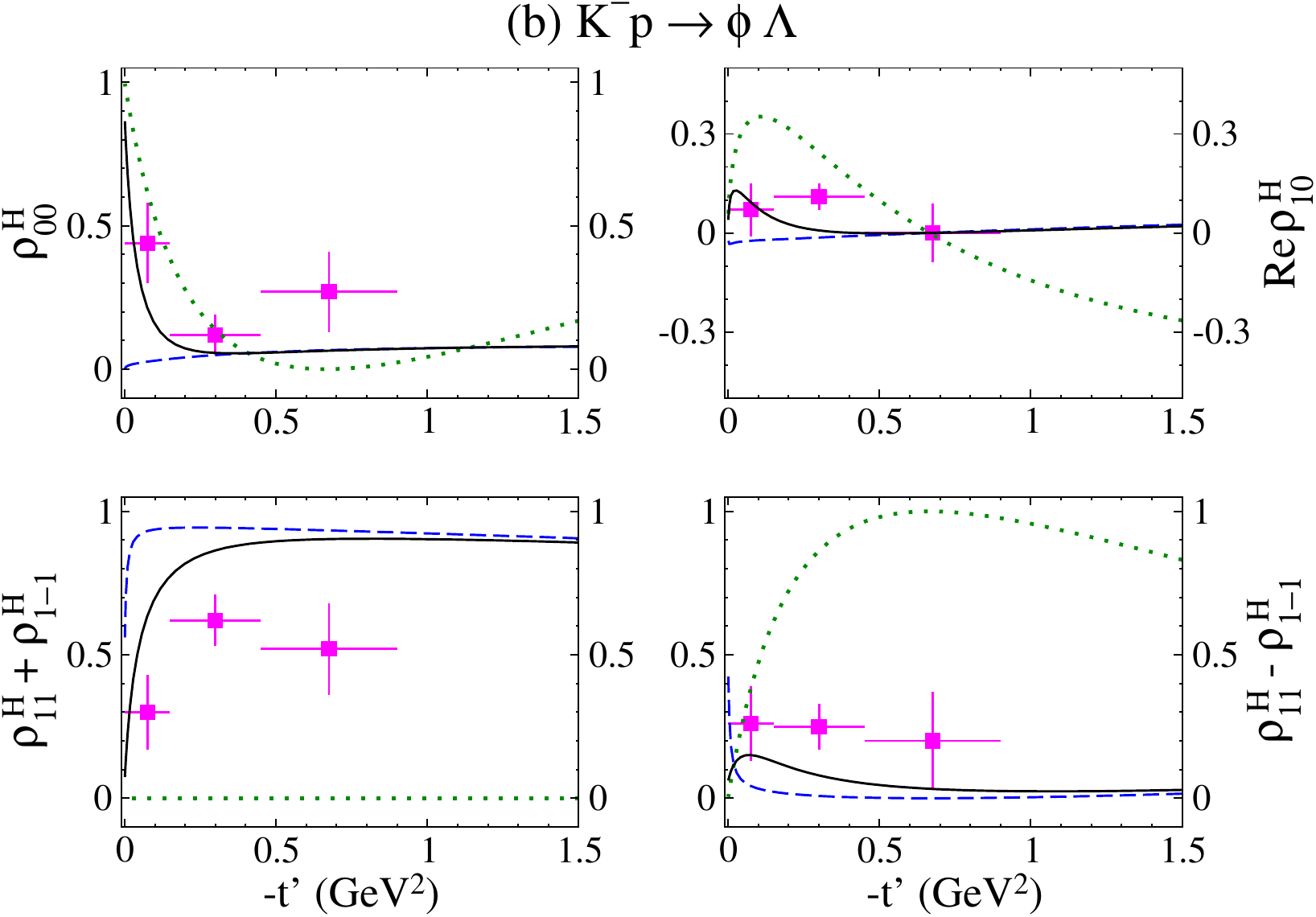}
\caption{SDMEs for the $K^- p \to \phi \Lambda$ reaction as functions of $-t'$.
(a) $\rho_{00}$, ${\rm Re},\rho_{10}$, and $\rho_{1-1}$ in the helicity and GJ frames
at $P_{\rm Lab}=4.2$ GeV.
(b) $\rho_{00}$, ${\rm Re},\rho_{10}$, and $\rho_{11} \pm \rho_{1-1}$ in the helicity
frame at $P_{\rm Lab}=5$ GeV.
Experimental data are taken from Refs.~\cite{Aguilar-Benitez:1972ngz} (circles)
and \cite{Ayres:1974aj} (squares).
Curve notations are the same as in Fig.~\ref{FIG05}.}
\label{FIG08}
\end{figure*}
The role of the individual contributions becomes evident in Fig.~\ref{FIG08}, where
the SDMEs are shown as functions of $-t'$ at $P_{\rm Lab} = 4.2$ and $5$ GeV in
both the helicity and GJ frames~\cite{Kim:2017hhm}.
In general, including the $K$-Reggeon exchange in addition to the $K^*$ Reggeon
contribution improves the overall description of the SDMEs.
In particular, the $t$ dependences of $\rho_{00}^H$ and ${\rm Re}\,\rho_{10}^H$ from
the $K$-Reggeon exchange closely follow those of the corresponding experimental
data.

\subsection{Charm production: $K^- p \to D_s^{*-} \Lambda_c^+,\,J/\psi \Lambda$}
\label{Sec:III-2}

\begin{figure*}[ht]
\centering
\includegraphics[width=7.0cm]{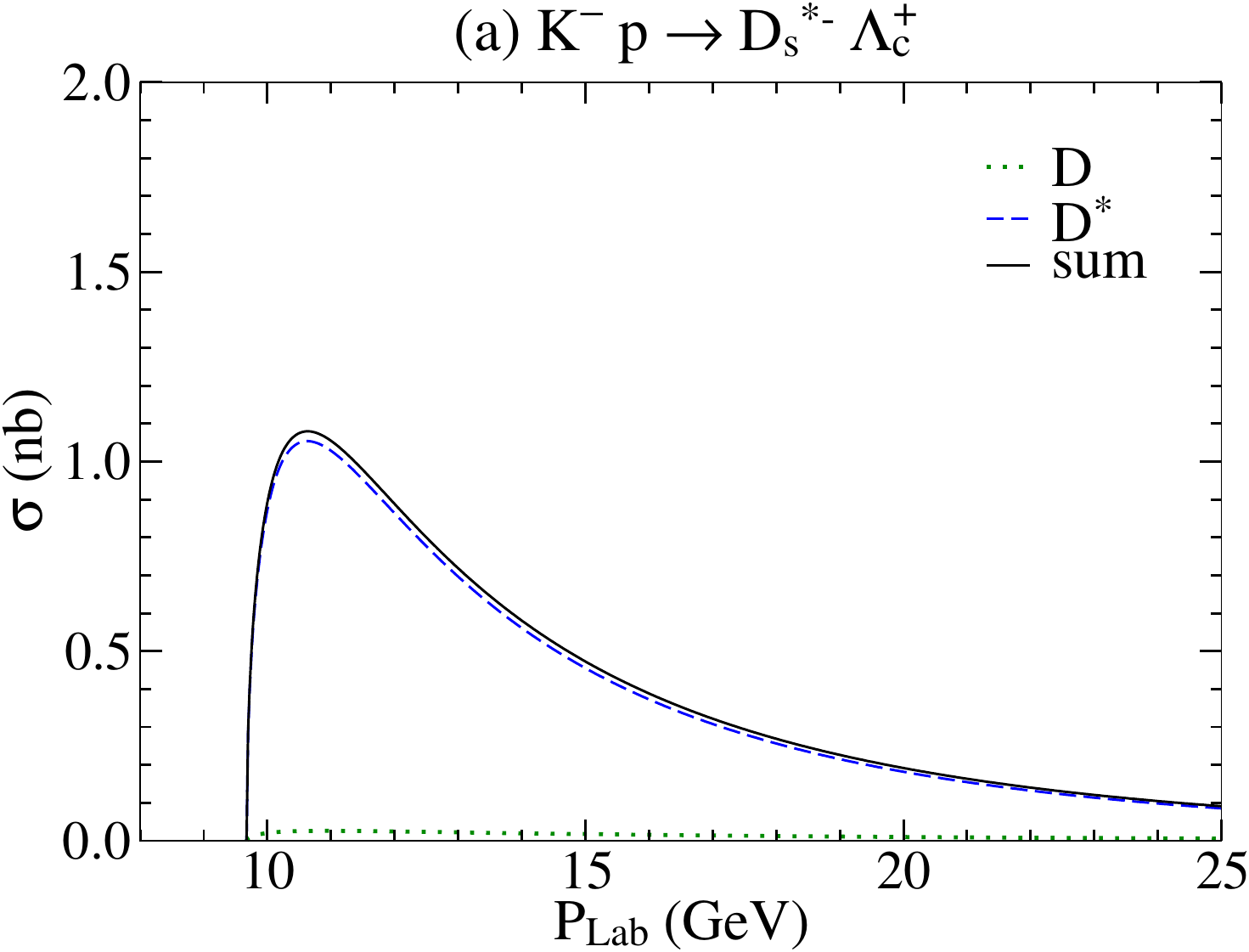} \hspace{1.5em}
\includegraphics[width=7.0cm]{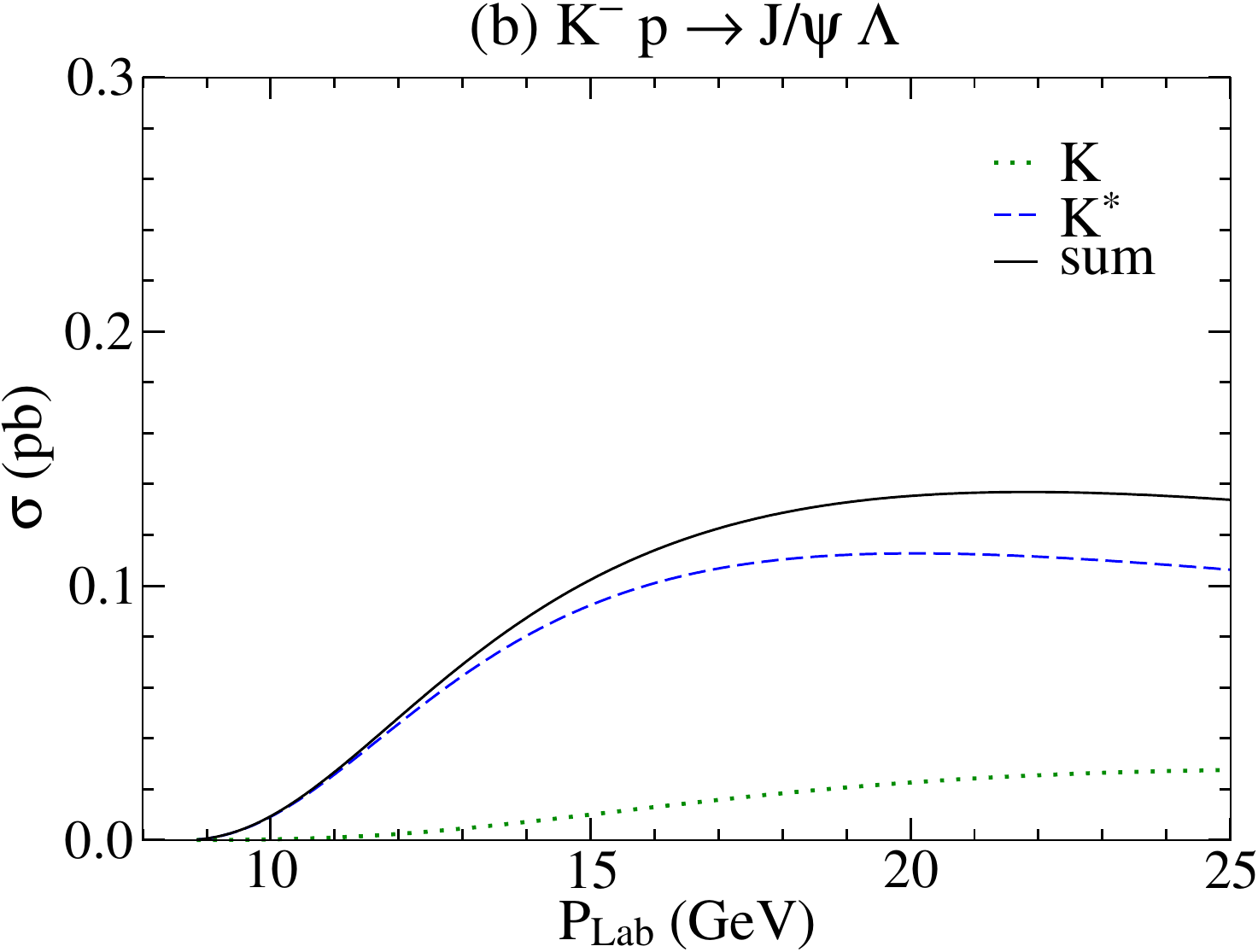}
\caption{Total cross sections for the (a) $K^- p \to D_s^{*-} \Lambda_c^+$ and (b)
$K^- p \to J/\psi \Lambda$ reactions as functions of $P_{\rm Lab}$.
The green dotted and blue dashed curves denote the contributions from pseudoscalar-
and vector-Reggeon exchanges, respectively, while the black solid curve represents
their sum.}
\label{FIG09}
\end{figure*}
We now turn to charm production described in Sec.~\ref{Sec:II-2} and compare the
results with those for the strangeness production $K^- p \to \phi \Lambda$.
Figure~\ref{FIG09} depicts the predicted total cross sections for the $K^- p \to
D_s^{*-} \Lambda_c^+$ and $K^- p \to J/\psi \Lambda$ reactions as functions of
$P_{\rm Lab}$.
In both cases, vector-meson Reggeon exchanges dominate, similarly to the
strangeness production channel.
Owing to the much larger mass gap in the final state of the $K^- p \to J/\psi
\Lambda$ channel, its total cross section increases more gradually with $P_{\rm Lab}$
than those of the $K^- p \to \phi \Lambda$ and $K^- p \to D_s^{*-} \Lambda_c^+$
reactions.
Our result for $K^- p \to J/\psi \Lambda$ is approximately three orders of magnitude
smaller than that predicted in Ref.~\cite{Clymton:2021thh}.

\begin{figure*}[ht]
\centering
\includegraphics[width=7.0cm]{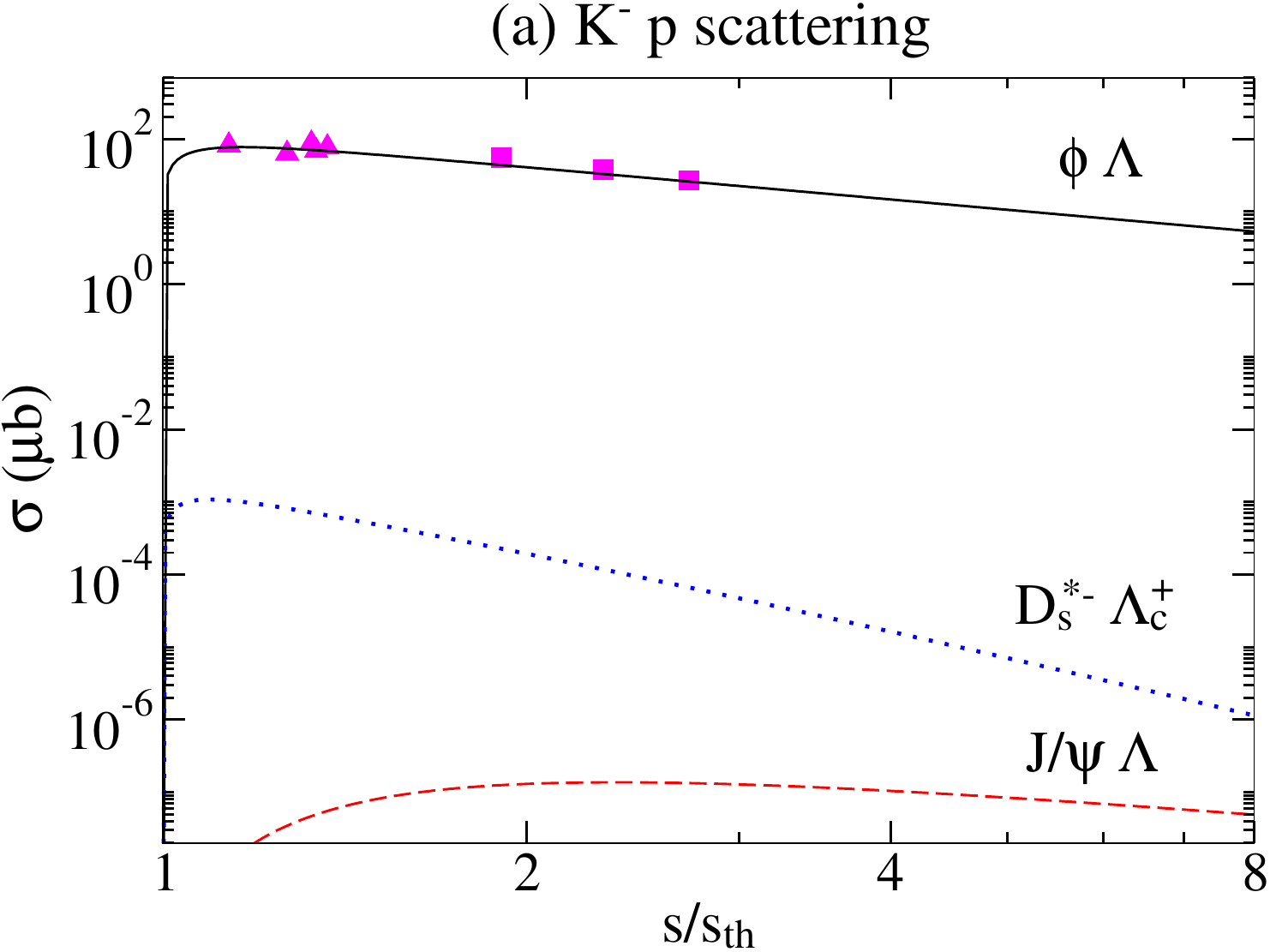} \hspace{1.5em}
\includegraphics[width=7.0cm]{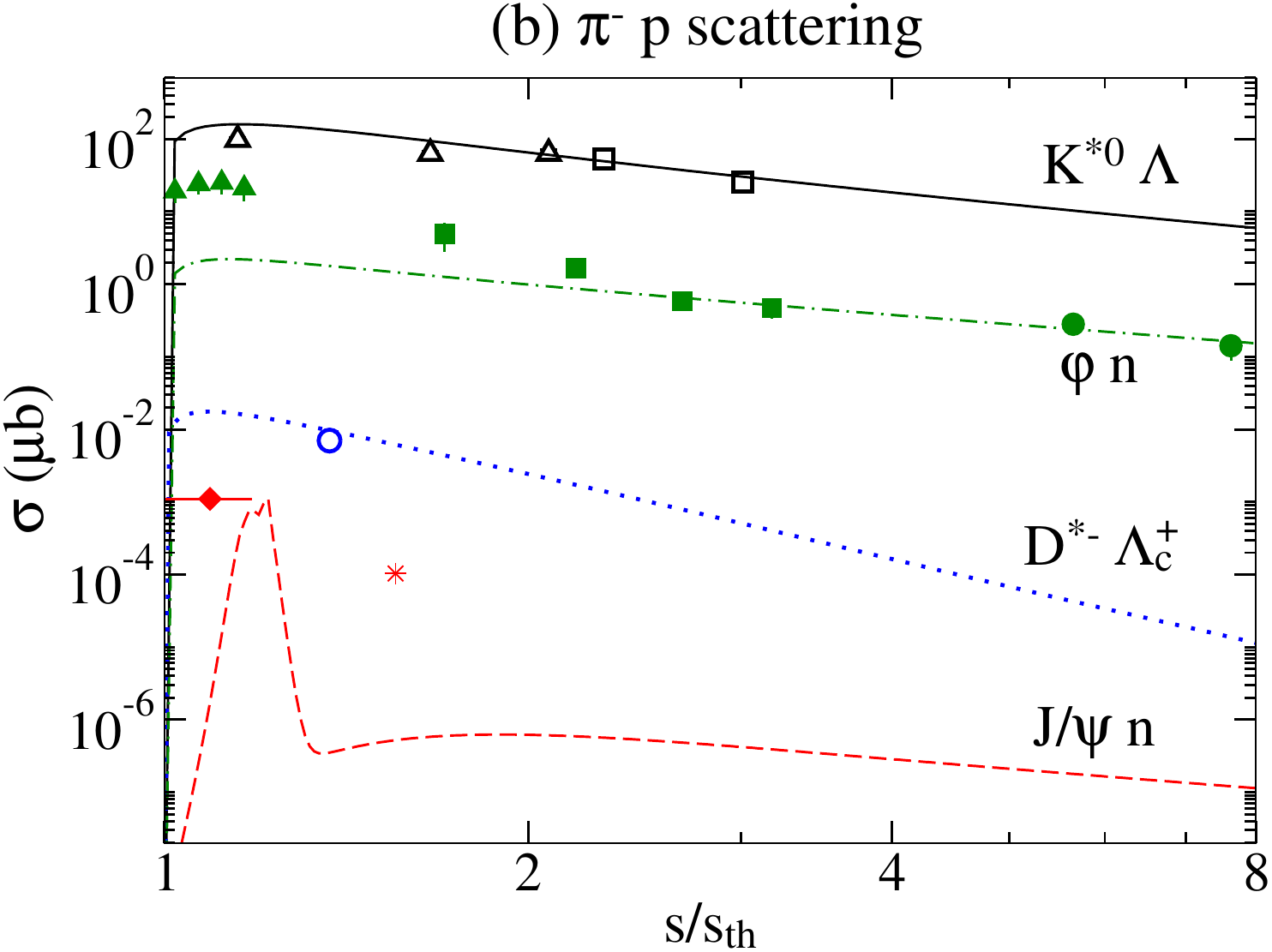}
\caption{(a) Total cross sections for the reactions $K^- p \to (\phi \Lambda$,
$D_s^{*-} \Lambda_c^+$, $J/\psi \Lambda)$ as functions of $s/s_{\rm th}$.
Experimental data are taken from Refs.~\cite{Lindsey:1966zz} (triangles) and
\cite{Ayres:1974aj} (squares).
(b) Total cross sections for the reactions $\pi^- p \to (K^{*0} \Lambda$, $\phi n$,
$D^{*-} \Lambda_c^+$, $J/\psi n)$.
Experimental data are taken from Refs.~\cite{Dahl:1967pg} (open triangles),
\cite{Crennell:1972km} (open squares), \cite{Courant:1977rk} (triangles), \cite{
Ayres:1974aj} (squares and circles), \cite{Christenson:1985ms} (open circles), \cite{
Jenkins:1977xb} (diamonds), and \cite{Chiang:1986gn} (stars).
Theoretical results are taken from Refs.~\cite{Kim:2015ita,Kim:2016cxr}.}
\label{FIG10}
\end{figure*}
A direct comparison among the $\phi \Lambda$, $D_s^{*-} \Lambda_c^+$, and
$J/\psi \Lambda$ production is given in Fig.~\ref{FIG10}(a) as functions of
$s/s_{\rm th}$, where $s_{\rm th}$ is the threshold energy of the corresponding
reaction.
The total cross section for the open-charm reaction $K^- p \to D_s^{*-} \Lambda_c^+$
is suppressed by approximately $5$-$6$ orders of magnitude compared with that for
the strangeness reaction $K^- p \to \phi \Lambda$ depending on the kinematical
region.
This suppression can be attributed mainly to the larger energy-scale parameter in
the charm sector, $s_{D (D^*)}^{K N : D_s^* \Lambda_c} > s_{K (K^*)}^{K N : \phi \Lambda}$, as
well as to the smaller coupling strength, $g_{D_s^* D^* K} < g_{\phi K^* K}$.
In the case of the hidden-charm reaction $K^- p \to J/\psi \Lambda$, the total cross
section is further suppressed by about $8$-$9$ orders of magnitude compared to that
for the strangeness reaction $K^- p \to \phi \Lambda$, even though the same Regge
parameters are employed.
This stronger suppression mainly arises from the much smaller coupling strength in
the charm sector, $g_{\phi K^* K} \simeq 3.5 \cdot 10^3 g_{J/\psi K^* K}$.

It is worthwhile to compare the results shown in Fig.~\ref{FIG10}(a) with those of
the relevant pion-induced reactions studied in Refs.~\cite{Kim:2015ita,Kim:2016cxr}
within a similar theoretical framework.
In Fig.~\ref{FIG10}(b), we compare the total cross sections for the reactions
$\pi^- p \to K^{*0} \Lambda$~\cite{Dahl:1967pg,Crennell:1972km},
$\phi n$~\cite{Ayres:1974aj,Courant:1977rk}, $D^{*-} \Lambda_c^+$~\cite{
Christenson:1985ms}, and $J/\psi n$~\cite{Jenkins:1977xb,Chiang:1986gn}.
The hidden-hadron production channels ($\phi n$, $J/\psi n$) are found to be more
suppressed than the corresponding open-hadron production channels ($K^{*0} \Lambda $,
$D^{*-} \Lambda_c^+$) due to the the Okubo-Zweig-Iizuka (OZI) rule~\cite{Okubo:1963fa,
Okubo:1963fa2,Okubo:1963fa3}, similar to the kaon-induced reactions shown in
Fig.~\ref{FIG10}(a).
The predicted cross section for the $D^{*-} \Lambda_c^+$ channel is in good agreement
with the available experimental upper limit~\cite{Christenson:1985ms}.
The peak structure observed in the $\pi^- p \to J/\psi n$ reaction originates from
the contributions of the hidden-charm pentaquark states $P_c (4380)$ and
$P_c (4450)$~\cite{LHCb:2015yax}.
The resulting cross section near threshold is close to the experimental upper
limit~\cite{Jenkins:1977xb,Chiang:1986gn}.
Note that the result for the $J/\psi n$ production should be regarded as a lower
bound on the cross section, since only the summed branching ratio ${\mathcal B}
(\phi \to \rho\pi + \pi^+\pi^-\pi^0) = 15.32\,\%$ is experimentally
available~\cite{PDG:2024cfk},
This leads to an ambiguity in determining the individual ${\mathcal B} (\phi \to
\rho\pi)$ channel.

\begin{figure*}[ht]
\centering
\includegraphics[width=8.5cm]{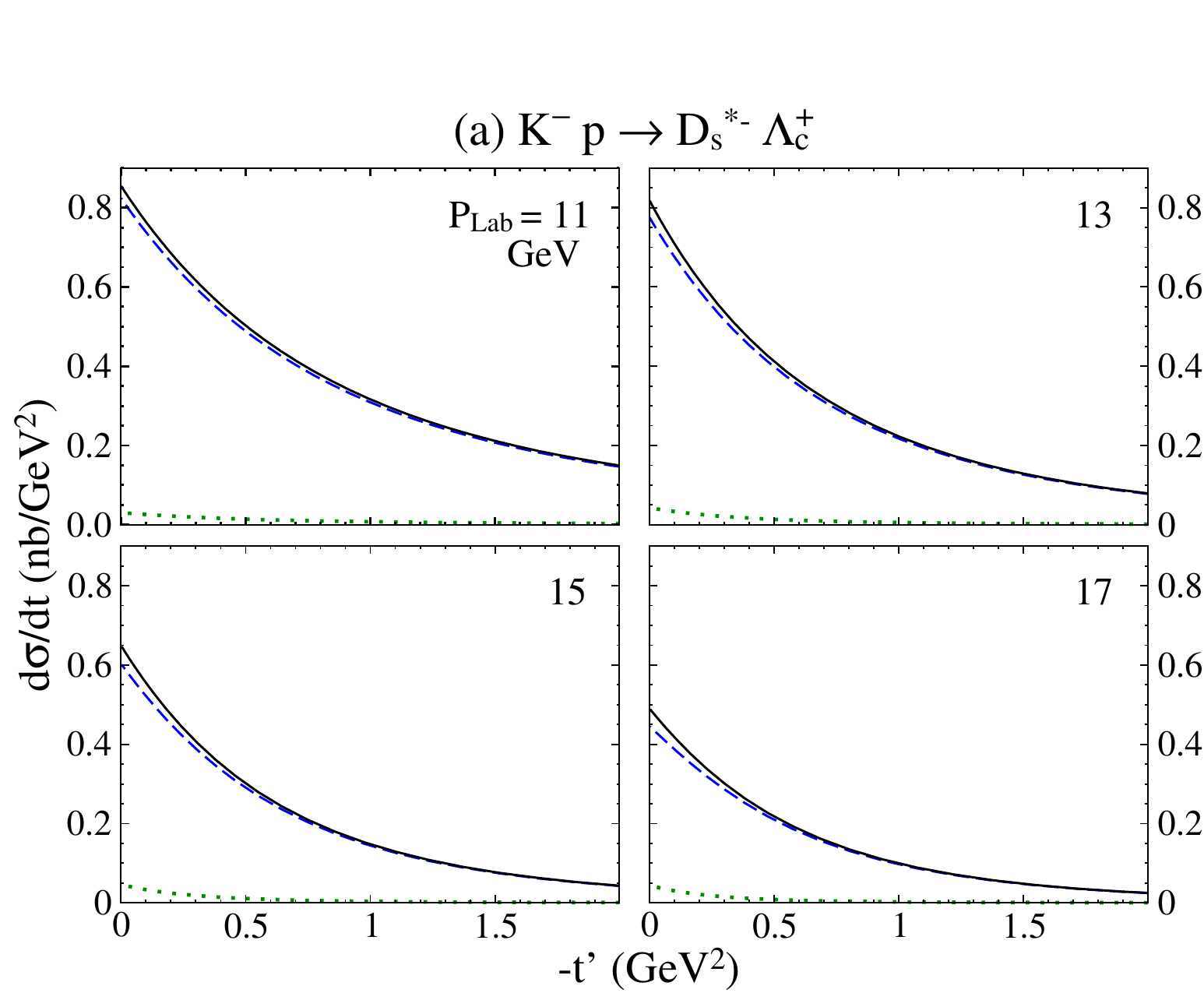} \hspace{1.5em}
\includegraphics[width=8.5cm]{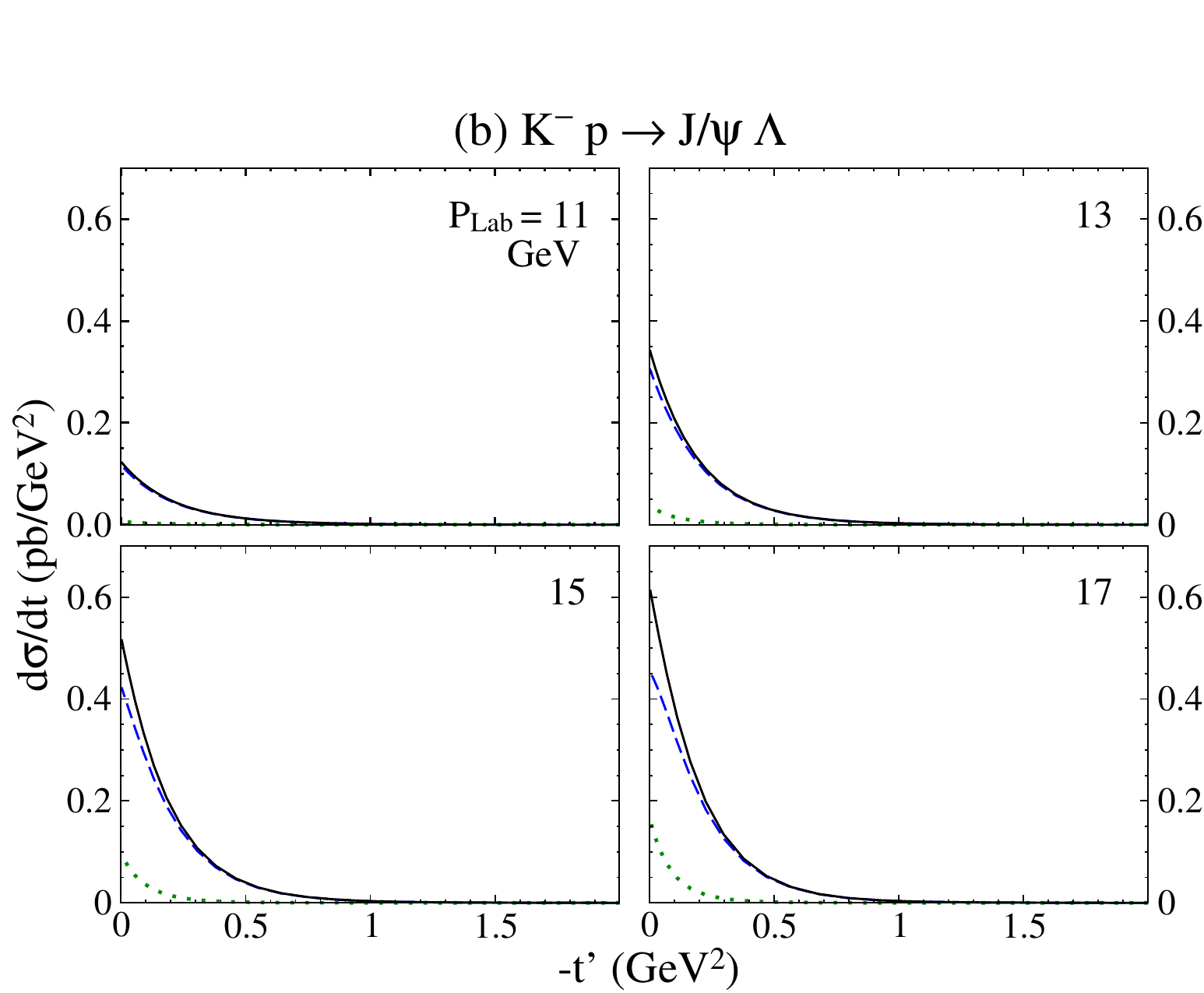}
\caption{$t$-dependent differential cross sections for the (a) $K^- p \to D_s^{*-}
\Lambda_c^+$ and (b) $K^- p \to J/\psi \Lambda$ reactions at four fixed beam energies.
Curve notations are the same as in Fig.~\ref{FIG09}.}
\label{FIG11}
\end{figure*}
Figure~\ref{FIG11} displays the predicted $t$-dependent differential cross sections
for the $K^- p \to D_s^{*-} \Lambda_c^+$ and $K^- p \to J/\psi \Lambda$ reactions at
four fixed beam energies.
As expected, forward peaks are observed in both cases, particularly in the latter
case.
The differential cross sections are expressed in units of nb/GeV$^2$ and pb/GeV$^2$,
respectively.
Consequently, future high-precision accelerator facilities will be required to
measure such small cross sections, particularly for the latter reaction.

\begin{figure*}[ht]
\centering
\includegraphics[width=16.0cm]{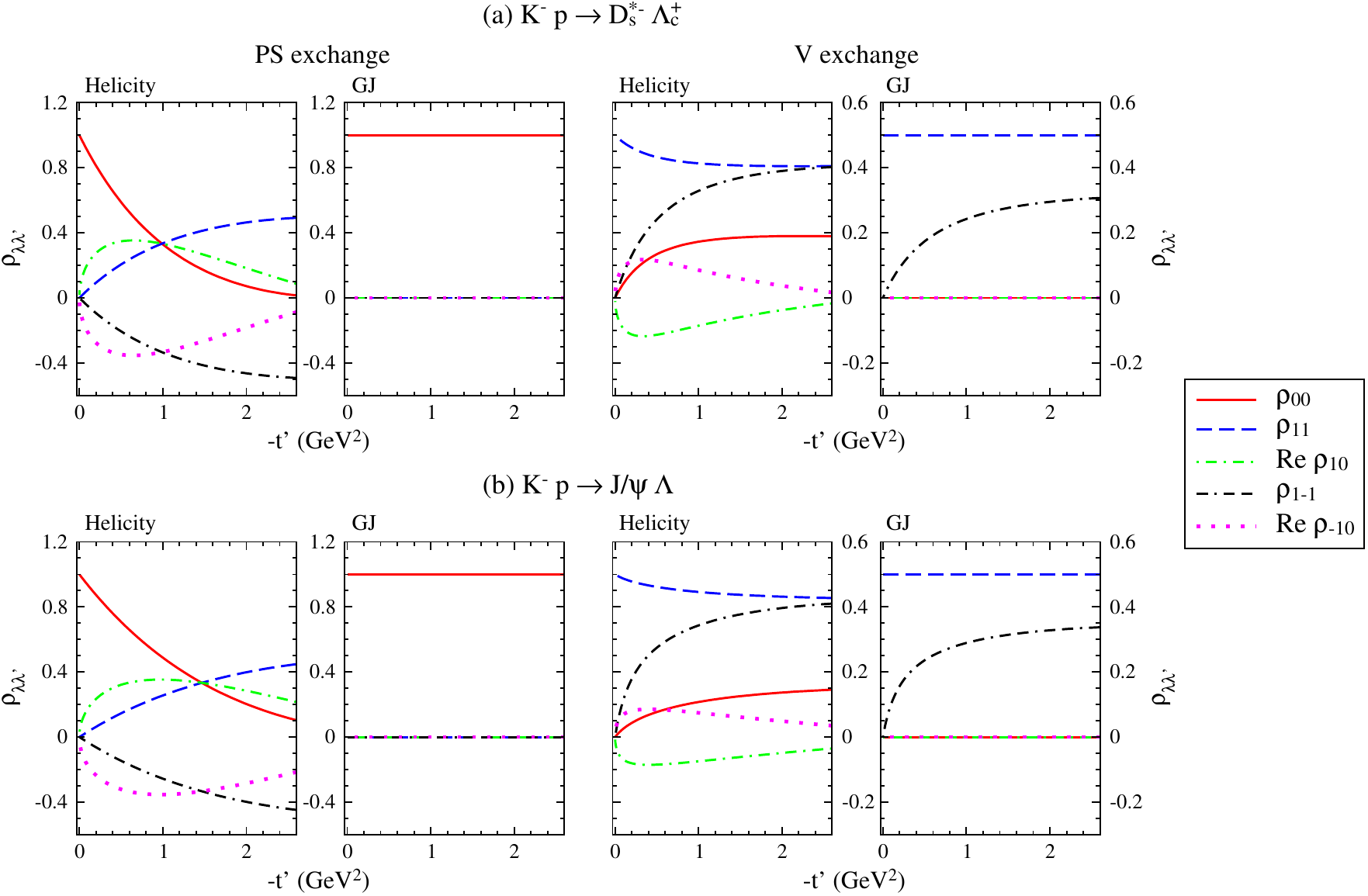}
\caption{SDMEs for the (a) $K^- p \to D_s^{*-} \Lambda_c^+$ and (b) $K^- p \to J/\psi
\Lambda$ reactions as functions of $-t'$ at $P_{\rm Lab}=15$ GeV.
The results for pseudoscalar- and vector-Reggeon exchanges are shown in the left
and right panels, respectively, in both the helicity and GJ frames.}
\label{FIG12}
\end{figure*}
\begin{figure*}[ht]
\centering
\includegraphics[width=15.0cm]{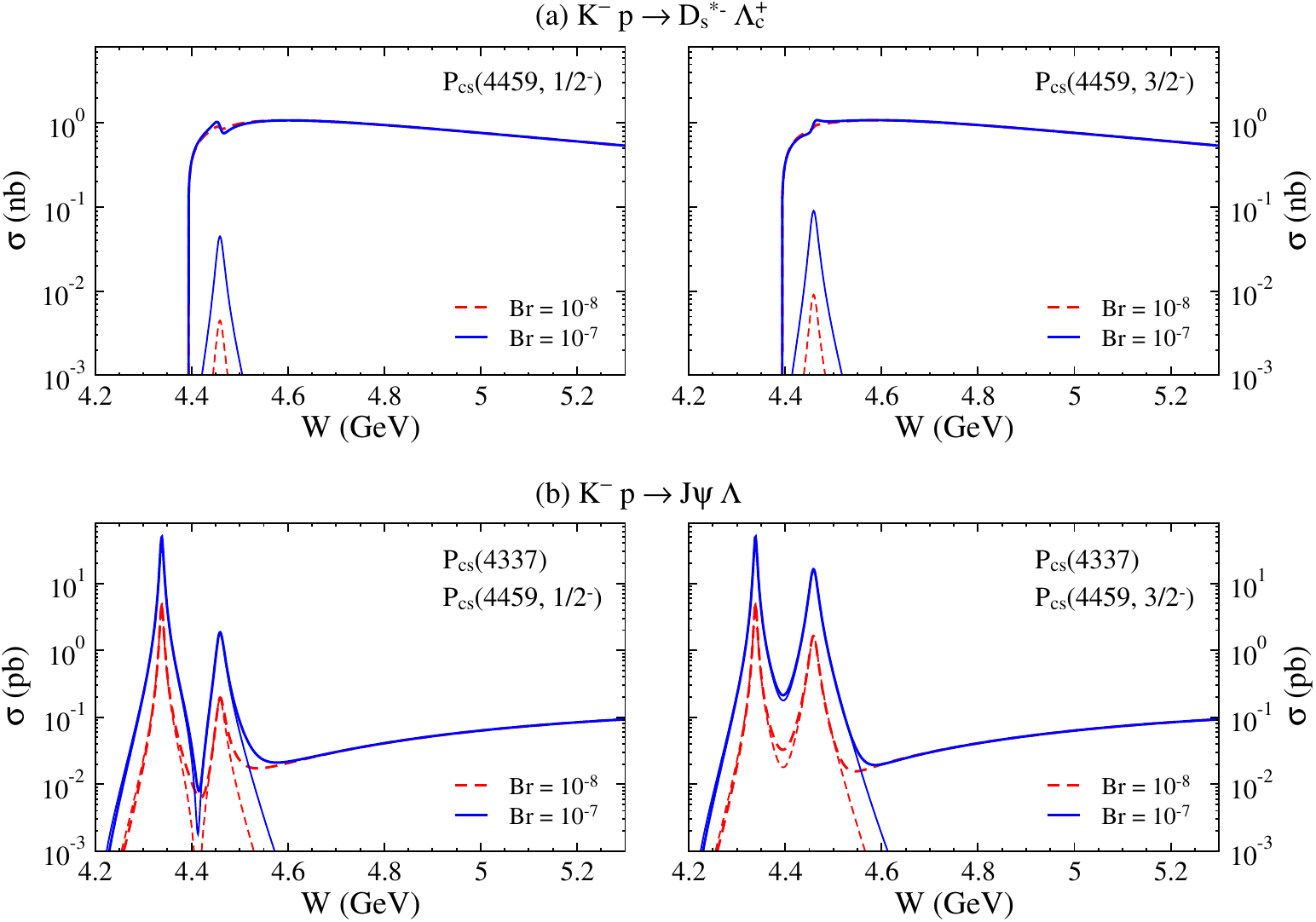}
\caption{Total cross sections for the (a) $K^- p \to D_s^{*-} \Lambda_c^+$ and (b)
$K^- p \to J/\psi \Lambda$ reactions as functions of $W$, including contributions
from the $P_{cs}(4337)^0$ and $P_{cs}(4459)^0$ states.
Left and right panels correspond to $J^P = 1/2^-$ and $3/2^-$ for $P_{cs}(4459)^0$,
respectively.
The branching ratio ${\mathcal B}(P_{cs} \to {\bar K} N)$ is taken to be $10^{-7}$ and
$10^{-8}$.}
\label{FIG13}
\end{figure*}
In Fig.~\ref{FIG12}, we show the individual predictions for the SDMEs as functions of
$-t'$ at $P_{\rm Lab} = 15$ GeV in both the helicity and GJ frames.
In principle, for the vector-meson Reggeon exchange, the matrix elements
$\rho_{\lambda\lambda'}$ with $|\lambda| = |\lambda'| = 1$ are specifically enhanced.
This behavior originates from the spin structure $\epsilon^{\mu\nu\alpha\beta}
\varepsilon^*_\mu (\lambda_V) k_{2\alpha} k_{1\beta}$ of the amplitude in
Eq.~(\ref{eq:Ampl1}).
In the vector meson rest frame, where $\bm{k_2} = (M_V,0,0,0)$ and $\bm{k_1}$
corresponds to the incoming kaon three-momentum $\bm{p_K}$,
this factor is proportional to the vector product ${\bm{\varepsilon}}^* (\lambda_V)
\times \bm{p}_K$.
In the helicity frame and at small momentum transfers, $\bm{p}_K$ has a large $z$
component and a small transverse component, which yields ${\bm\varepsilon}^*
(\lambda_V) \times \bm{p}_K \simeq i\lambda_V {\bm\varepsilon}^*(\lambda_V)
|\bm{p}_K|$, thereby leading to a pronounced enhancement of
$\rho_{|\lambda| = 1,\, |\lambda^\prime| = 1}$.
In the GJ frame, $\bm{p}_K$ is aligned with the quantization axis;
consequently, $\rho_{\lambda\lambda'}$ vanishes whenever either $\lambda = 0$ or
$\lambda' = 0$.
For the pseudoscalar-Reggeon exchange, however, the situation is markedly
different.
In this case, the scattering amplitude is proportional to the scalar product
$\bm{\varepsilon}^*(\lambda_V) \cdot \bm{p}_K$ in Eq.~(\ref{eq:Ampl1}), which
strongly enhances $\rho_{00}$ in the GJ frame, so that $\rho_{00}=1$, while all other
$\rho_{\lambda\lambda'}$ vanish~\cite{Kim:2017hhm}.

\subsection{Pentaquark contributions to $K^- p \to
D_s^{*-} \Lambda_c^+,\,J/\psi \Lambda$}
\label{Sec:III-3}

Since we obtained reasonable estimates for the background contributions to the $K^- p
\to D_s^{*-}\Lambda_c^+$ and $K^- p \to J/\psi \Lambda$ reactions in the previous
subsection, we now investigate the $s$-channel $P_{cs}$ contributions illustrated in
Fig.~\ref{FIG04}.
Using the branching ratios ${\mathcal B}(P_{cs} \to \bar D_s^* \Lambda_c)$ and
${\mathcal B}(P_{cs} \to J/\psi \Lambda)$ determined in Eqs.~(\ref{eq:BR_Pcs_1}) and
(\ref{eq:BR_Pcs_2}), we take the branching ratios for $P_{cs} \to {\bar K} N$ to be
$10^{-7}$ and $10^{-8}$, treating them as free parameters.
The predicted total cross sections are shown in Fig.~\ref{FIG13} as functions of the
c.m. energy $W = \sqrt{s} $, where the results for $P_{cs}(4459)^0$ with $J^P = 1/2^-$
and $3/2^-$ are displayed in the left and right panels, respectively.
The contribution from the $P_{cs}$ state with $J^P = 3/2^-$ is larger than that with
$J^P = 1/2^-$ for both reactions.
Since the background contribution to the (a) $K^- p \to D_s^{*-} \Lambda_c^+$ reaction
is much larger than that to the (b) $K^- p \to J/\psi \Lambda$ reaction, the
$P_{cs}(4459)^0$ signal is difficult to observe in the $D_s^{*-} \Lambda_c^+$ channel.
In contrast, both $P_{cs}(4337)^0$ and $P_{cs}(4459)^0$ give sizable contributions to
$J/\psi \Lambda$ production, with the contribution from $P_{cs}(4337)^0$ being larger.

\section{Summary and Conclusion}
\label{Sec:IV}

In this work, we investigated strangeness production in the $K^- p \to \phi \Lambda$
reaction within a hybrid Regge framework.
The reaction dynamics are found to be governed predominantly by $t$-channel
exchanges.
In particular, the $K^*$-Reggeon exchange provides the leading contribution over the
considered energy range, while the $K$-Reggeon exchange plays a nonnegligible role,
especially in reproducing the observed SDMEs.
Contributions from the $s$-channel $\Lambda$ and $u$-channel nucleon exchanges turn
out to be strongly suppressed.

Extending the same Regge-based approach, we analyzed the open-charm $K^- p \to
D_s^{*-} \Lambda_c^+$ and hidden-charm $K^- p \to J/\psi \Lambda$ reactions within a
QGSM-motivated framework.
The Regge trajectories $\alpha(t)$ and energy-scale parameters $s_0$ were
determined consistently, thereby reducing model dependence and theoretical
uncertainties.
Our results indicate that the production rates for the open- and hidden-charm
channels are strongly suppressed, by roughly 5–6 and 8–9 orders of magnitude,
respectively, compared with the strangeness-production channel, depending on the
kinematical region.
This strong suppression originates mainly from the larger energy-scale parameters
and much smaller effective coupling strengths in the charm sector.

We further explored possible $s$-channel contributions from the hidden-charm
pentaquark states with strangeness, $P_{cs}(4337)^0$ and $P_{cs}(4459)^0$.
While their effects are difficult to isolate in the open-charm channel due to the
large nonresonant background, they may produce noticeable enhancements in the
$J/\psi \Lambda$ channel, suggesting that hidden-charm production could provide a
more favorable environment for studying the $P_{cs}$ states.

Other reaction mechanisms, such as scalar $\kappa$ and axial-vector $K_1$ exchanges,
are in principle possible for the $K^- p \to \phi \Lambda$ reaction but were not
included in the present work, since their contributions are expected to be small
compared with those of the $K$- and $K^*$-Reggeon exchanges.
In addition, the tetraquark candidate $Z_{cs}(4000)^+$, observed in the $J/\psi K^+$
channel and having a possible $c\bar{c}u\bar{s}$ quark content~\cite{LHCb:2021uow},
could contribute to the $K^- p \to J/\psi \Lambda$ reaction via $t$-channel
exchange.
Such effects were also not considered here and deserve further study.

Our predictions for charm-production observables, including total and differential
cross sections as well as SDMEs, can serve as useful benchmarks for future
measurements at J-PARC~\cite{Aoki:2021cqa,Takahashi:2012cka,Agari:2012gj}.
Within the same theoretical framework, the present study can be extended to other
strangeness and charm production processes in kaon-induced reactions, such as $K^- p
\to \phi \Sigma^0$, $D_s^{*-} \Sigma_c^+$, $J/\psi \Sigma^0$, as well as to
pion-induced reactions like $\pi^- p \to K^{*0} \Sigma^0$, $D^{*0} \Sigma_c^0$.
The positive-parity states $D_{s0}(2317)$ and $D_{s1}(2460)$ lie about 40 MeV below
the $DK$ and $D^* K$ thresholds, respectively, and are widely discussed in
connection with $DK$ and $D^* K$ molecular interpretations~\cite{Liu:2012zya,
Faessler:2007us}.
It is therefore interesting to investigate their production mechanisms in the
$K^- p \to D_{s0}(2317) \Lambda_c^+$ and $K^- p \to D_{s1}(2460) \Lambda_c^+$
reactions~\cite{Zhu:2019vnr}.
Such studies may provide further insight into their internal structure and
underlying dynamics.
Working along these lines is currently in progress.

\section*{Acknowledgments}

The work was supported by the Basic Science Research Program through the National
Research Foundation of Korea (NRF) under Grants No. RS-2021-NR060129.



\end{document}